\documentclass[aps,reprint, prx, superscriptaddress, showpacs,nobibnotes]{revtex4-1}
\usepackage{graphicx}
\usepackage{bm,amsmath}
\usepackage{systeme}
\usepackage{hyperref}
\usepackage{mathrsfs}
\usepackage{dcolumn}
\usepackage{natbib}
\usepackage{braket}
\usepackage[dvipsnames]{xcolor}
\usepackage{ulem}
\usepackage[version=4]{mhchem}
\usepackage[title]{appendix}
\usepackage{blkarray}

\begin{document}
\title{Slow magnetic response and associated spin-lattice relaxation dynamics in rare-earth vanadates revealed by AC susceptibility and magnetocaloric methods}

\author{Yuntian Li}
\affiliation{Department of Applied Physics, Stanford University, Stanford, California 94305, USA}
\affiliation{Geballe Laboratory for Advanced Materials, Stanford University, Stanford, California 94305, USA}
\author{Linda Ye}
\affiliation{Department of Applied Physics, Stanford University, Stanford, California 94305, USA}
\affiliation{Geballe Laboratory for Advanced Materials, Stanford University, Stanford, California 94305, USA}
\author{Mark P. Zic}
\affiliation{Geballe Laboratory for Advanced Materials, Stanford University, Stanford, California 94305, USA}
\affiliation{Department of Physics, Stanford University, Stanford, California 94305, USA}

\author{Arkady Shekhter}
\affiliation{Los Alamos National Laboratory, Los Alamos, New Mexico 87545, USA}
 
\author{Ian R. Fisher}
\affiliation{Department of Applied Physics, Stanford University, Stanford, California 94305, USA}
\affiliation{Geballe Laboratory for Advanced Materials, Stanford University, Stanford, California 94305, USA}

\date{\today}

\begin{abstract}

This report presents a new technique to probe the quantitative dynamical response of the magnetic field induced heating/cooling process in rare-earth vanadium materials. The approach combines AC magnetic susceptibility and AC caloric measurements to reveal the intrinsic timescale associated with the magnetic relaxation process of rare-earth ions at low temperatures. Utilizing the well-known crystal field levels in \ce{YbVO4}, we prove and demonstrate a discretized thermal analysis through a common spin-lattice relaxation phenomenon. The demonstration experiment presented in this study provides a general approach to quantitatively address multiple measured quantities in one unified discretized thermal circuit analysis. It can be extended to study other magnetic, dielectric, and elastic materials exhibiting a complex response to an external driving field in the presence of intrinsic interactions and fluctuations, particularly when an energy dissipation process is within an accessible frequency regime.

\end{abstract}

\maketitle

\section{Introduction}

\begin{figure}[ht!]
	\includegraphics[width = \columnwidth]{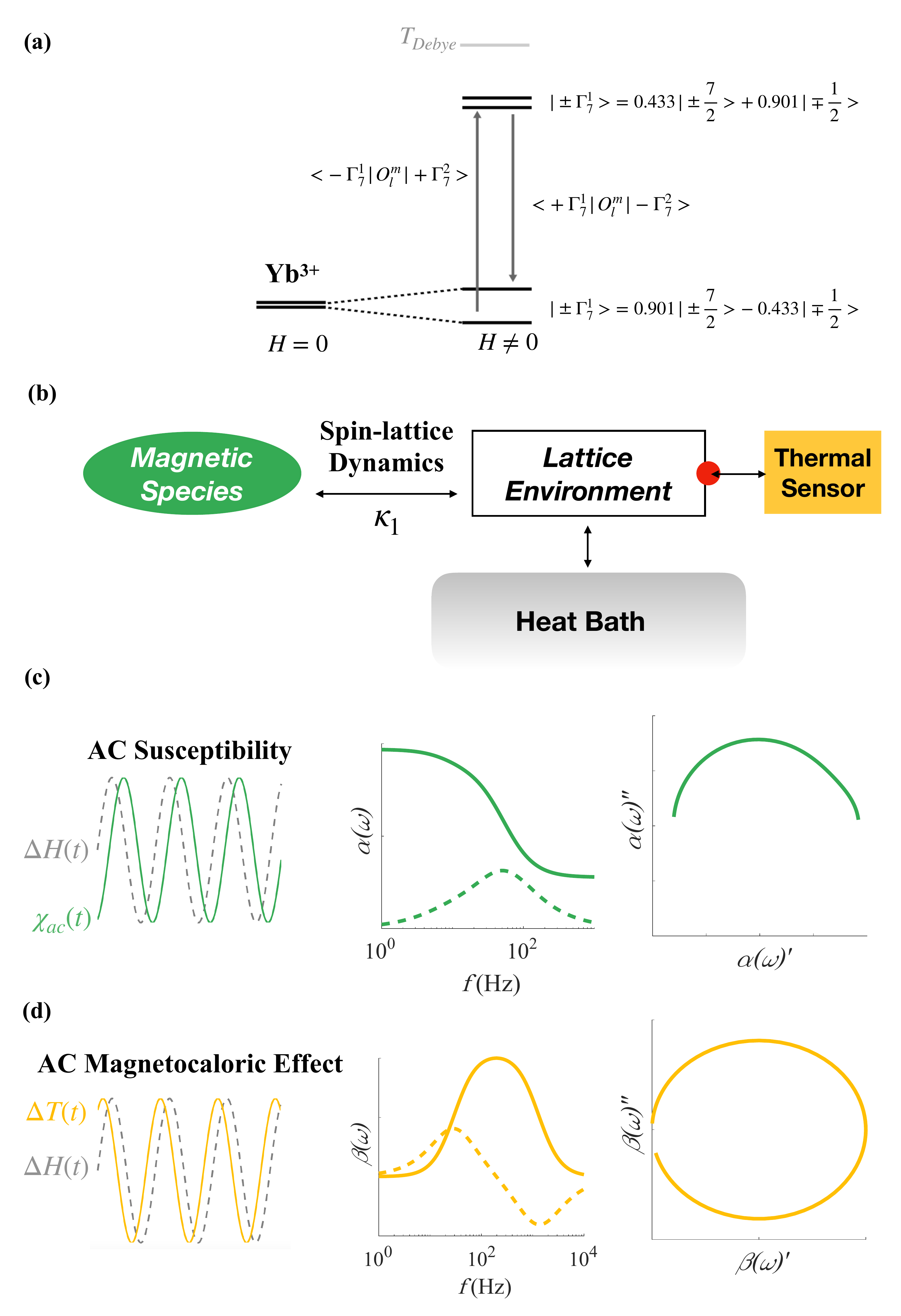}
	\caption{\label{fig-1} (a) Schematics of crystal electric field levels and indirect transition corresponding to \ce{Yb^{3+}} ion embedded in the \ce{RVO4} crystal field environment. The lower two levels represent Kramers' ground-state doublets, which undergo splitting when an external magnetic field is along the c-axis. The higher states correspond to the second excited states of the crystal field levels. (b) Discretized spin-lattice thermal circuit model components that are summarized from the experimental observations, highlighting the major time constant measured by frequency-dependent AC techniques caused by the indirect transition of the ground-state doublets. (c,d) Solution of the equivalent circuit of systems based on the internal discretized thermal circuit model. In the left panels, the response signal is plotted to illustrate the phase difference between the time-dependent solution and the signals. In the middle panels, solid lines represent the real components ($\alpha'$ and $\beta'$), and dashed lines represent the imaginary components ($\alpha''$ and $\beta''$). In the right panels, the real components of each solution are plotted against their respective imaginary parts in a Cole-Cole plot. This representation maintains a fixed aspect ratio of 1:1 for the $x$ and $y$ axes.}
\end{figure}


AC magnetic susceptibility describes the magnetic response of a material to an oscillating magnetic field\cite{Topping_2019}. The zero frequency limit is referred to as the isothermal susceptibility because all the internal subsystems (spins, phonons, etc) are in equilibrium. For non-zero frequencies, the spin and lattice subsystems may not have enough time to equilibrate and the effective response is not isothermal anymore, rendering the measured response frequency-dependent. A classic example is the phonon-bottleneck effect that is observed at low temperatures in many magnetic materials that exhibit magnetic anisotropy\cite{Orbach1961}. Oscillating magnetic fields also necessarily result in a caloric response due to changes in the entropy. The induced thermal relaxation to the bath, combined with the internal relaxation processes, is typically neglected in magnetic measurements \cite{Quilliam2008, Quilliam2011, Coca2014}. To account for magnitization-induced heat dynamics, a full treatment of the combined magnetic and thermal response is required. Here, we provide a framework to measure and describe the AC magnetic susceptibility and AC magnetocaloric response of materials.

We studied the magnetic insulator \ce{YbVO4} by associating the AC magnetic susceptibility measurement with its thermal response due to a significant magnetocaloric effect at low temperatures. The 4f spin degree of freedom in \ce{Yb} ion is unique in that it hosts a pseudospin 1/2 Kramers doublet that is weakly coupled to other \ce{Yb} ion states under the influence of perturbing fields (shown in Fig.\ref{fig-1}(a)). The effective exchange interaction of Kramers doublet on nearby ions is primarily via dipolar interactions and therefore is weak, as indicated by a small Neel temperature of 93 mK\cite{Radhakrishna1981}. Weak exchange as well as weak coupling to crystal fields make the dynamics of \ce{Yb} pseudospin very slow. This presents some challenges for conventional measurements of the magnetic response. In particular, one must account for the accompanying non-adiabatic heat exchange between Yb pseudospin and the lattice, which necessarily accompanies the magnetization change in an applied magnetic field. Such thermal exchange alters the observed magnetic response of an experimental setup that is typically used to measure magnetization.

Under an applied magnetic field along the easy c-axis of the crystal direction, the Kramers ground state doublet degeneracy of the \ce{Yb} ions in \ce{YbVO4} is split. If this is done fast enough (i.e., adiabatically), the combined spin and lattice system is no longer in equilibrium and must relax. Spin-lattice relaxation occurs by transitions between the distinct Crystal Electric Field (CEF) magnetic states and is either direct (involving a single phonon) or indirect (the transition involves a third or more intermediate states and two or more phonons). For Yb ions in \ce{YbVO4}, direct transitions are not allowed by symmetry (see Appendix. \ref{app:CEF}), and the ion can only relax by indirect processes (Fig. \ref{fig-1}(a)). Zeeman splitting of the CEF eigenstates necessarily leads to a strong field dependence on the associated relaxation rates, making this an especially useful material system to identify internal relaxation effects. 

At a temperature much lower than the Debye temperature of the lattice, very few phonons are available to participate in the dynamics, resulting in a slow buildup of the magnetization to reach its isothermal value due to the low transition probability (i.e., a manifestation of the phonon bottleneck effect). The indirect nature of transition effectively decouples the material into a separate bath of spins and a bath of phonons, resulting in large internal time constants that describe the non-instantaneous relaxation behavior. The associated slow relaxation effects in rare-earth compounds that exhibit large magnetic anisotropies have been studied since the 1960s \cite{Orbach1961}. The well-defined magnetic properties of \ce{YbVO4} imply that a proposed thermal analysis approach can be confirmed experimentally in this system. 

The magnetocaloric (MCE) effect refers to the temperature change in a material that responds to variations in an external magnetic field \cite{Warburg1881, Kohama2010}. From a first-principle thermodynamic description, the MCE has been extensively studied and applied in various fields, such as adiabatic demagnetization refrigeration techniques \cite{Bruck_2005} and characterizing the order of phase transitions \cite{Law2018,Pereira2024}. The dominant method that is typically used to make MCE measurements is in the time domain, in response to a swept magnetic field (see, for example, \cite{Kohama2010}). While several previous studies attempted to measure the MCE in periodically changing fields (i.e. an AC, or dynamic, MCE) \cite{FISCHER199179, Tokiwa2011}, quantitative analysis of a thermal transfer function is usually limited by choice of materials and temperature regime \cite{ALIEV2016601, ALIEV2022169300}, and to date a formal analysis has not been available. 

Based on a large magnetic and MCE response associated with $4f$ spin entropy changes can be anticipated and experimentally obtained in the paramagnetic state \cite{Kaze2006, Palacios2018}, we perform a combination of $\chi^{ac}$ and AC MCE measurements as a set of reciprocal approaches to measure the internal time constant associated with the spin-lattice relaxation rate based on the thermodynamic response of the material. As shown in Fig. 1(b), the measured heat conduction rate $\kappa_l$ can be dynamically described by an effective spin-lattice energy exchange thanks to the indirect transition process between the ground state doublet and its well-separated excited states. $\kappa_l$ obtained from a caloric effect is different from the lattice thermal conductivity in a thermal transport or susceptibility measurement (see Appendix. \ref{app:Extrinsic_Kappa}). 

It is worth mention that common engineering methods, such as finite element simulations, computational fluid dynamics \cite{Tang24}, and thermal imaging \cite{Iguchi2023} primarily focus on characterizing heat flow based on well-established thermal constants and geometric effects. Such simulations require well-defined boundary conditions and experimentally determined constants. If, however, a material substance demonstrates a complex internal response and caloric effect, the efficacy of any simulation methods is significantly reduced.

In this article, we consider the typical magnetization measurement setup and show how it can be analyzed to extract the intrinsic properties of the \ce{Yb} spin subsystem. In particular, we find that the intrinsic magnetic response of the \ce{Yb} 4f spin subsystem in \ce{YbVO4} is nondispersive, at least in the frequency range of 1 kilohertz and smaller. The major finding of this report is summarized in Fig. \ref{fig-1}(c,d), where two specific response functions $\alpha(\omega)$ and $\beta(\omega)$ can be solved analytically from a physically equivalent thermal circuit model, and can be both measured experimentally.

\section{Experimental methods}

Single crystals of \ce{YbVO4} and \ce{GdVO4} were synthesized via slow cooling in a flux of \ce{Pb2V2O7} using a mixture of high-purity rare-earth oxides precursors, \ce{Yb2O3} ($99.99\%$ purity from Alfa Aesar, CAS Number: 1314-37-0) and \ce{Gd2O3} ($99.99\%$ purity from Alfa Aesar, CAS Number: 12064-62-9). More details related to the flux-growth synthesis method can be found in Refs. \cite{feigelson1968flux, smith1974flux, oka2006crystal}. 

AC susceptibility measurements were performed in a Quantum Design Magnetic Property Measurement System (MPMS) with the AC susceptometer measurement options. The technique and procedure of the measurement techniques can be found elsewhere \cite{Topping_2019}. 

AC MCE was performed using a customized probe within the same MPMS, ensuring a direct comparison of the two techniques can be made. The MCE measurement device consists of a polished sample with a thickness between 30-100 $\mu m$, a thermometer attached to the top surface of the sample, and a quartz sample holder that is thermally anchored to the bottom surface of the sample. The thermoresistor attached to the sample is connected with a Wheatstone resistance bridge that is anchored to the heat bath (located on the part of the probe far apart from the measured material). More details on the materials for the device fabrication, data analysis, heat transfer, and thermodynamic considerations can be found in the following text and Appendix sections.

\section{Heat exchange model}

Motivated by characterizing new magnetic materials that exhibit spin-lattice relaxations, we first introduce a heat exchange model based on a discretized thermal analog circuit approach to extend the analysis of the AC MCE into the frequency domain. Similar to a previous research that utilized AC thermal impedance as a non-resonant method to access spin-lattice relaxation dynamics \cite{Khansili2023}, this work extends such method and applies it to understanding complex caloric responses. The experimental setup includes intrinsically several characteristic times associated, at the very least,  with the heat exchange between the sample and the bath as well as slow heat exchange between the 4f and phonon subsystems. To account for these, we first describe the complete thermodynamic model of the experimental setup. 

Fig. \ref{fig-1}(b) shows the schematic heat flow diagram. We apply a periodic magnetic field $H(t)$ and measure the magnetization of the \ce{Yb} 4f spin subsystem $M(t)$ as well as the temperature of the lattice $T^{lat}(t)$. As we apply a magnetic field, the magnetization and temperature of the 4f subsystem change, and it is no longer in thermal equilibrium with the lattice. This ensues the non-adiabatic (dissipative) heat exchange between the 4f spin and the lattice. The thermodynamics of the 4f spin is generated by their free energy (using $T$ and $H$ as variables)
\begin{equation}\label{4f-free-energy}
\begin{aligned}
    dF^{4f} = -S^{4f}dT^{4f} - M^{4f}dH^{4f}
\end{aligned}
\end{equation}
where $S^{4f}$ and $M^{4f}$ are the 4f subsystem's entropy and magnetization. Taking the derivative of $dF^{4f}$ with respect to $T^{4f}$ and $H$, we could write a matrix of the thermodynamic coefficients that describes the changes in the entropy and magnetization\cite{Landau}:


\begin{equation}\label{4f_cross_relaxation}
\begin{aligned}
   \left[ \begin{array}{l}
    dS \\ dM 
    \end{array} \right]^{4f} 
    = 
 \left[   \begin{array}{cc}
     \frac{C}{T_0} & \gamma \\ \gamma & \chi 
    \end{array} \right]^{4f}
\left[     \begin{array}{l}
    dT \\ dH 
    \end{array} \right]^{4f} 
\end{aligned}
\end{equation}
where $C$, $\gamma$, and $\chi$ are specific heat, magnetocaloric, and magnetic susceptibility coefficients. 
If there were no heat exchange between the pseudospin subsystem and the lattice, there would be no entropy change in the pseudospin subsystem, $dS^{4f}=0$. Thus we need to consider non-adiabatic heat exchange between pseudospins  and lattice as well as lattice and heat bath, 
\begin{equation}\label{4f_heat_equation}
\begin{aligned} 
        q_{4f \leftarrow lat} &= T^{4f}\frac{dS^{4f} }{dt}  = 
    -\kappa_l (dT^{4f}-dT^{lat})  
        \end{aligned}
        \end{equation} 
where $q_{4f \leftarrow lat}$ is heat flux from lattice to the $4f$ spins and $\kappa_l$ is the heat link between the two. These equations need to be accompanied by equations for heat flow into the lattice, both from the pseudospins and from the heat bath:
\begin{equation}\label{lattice_heat_equation}
\begin{aligned} 
T^{lat}\frac{dS^{lat} }{dt} &= -  q_{4f \leftarrow lat} - q_{bat \leftarrow lat}  \\
&= \kappa_l (dT^{4f}-dT^{lat}) - \kappa_b (dT^{lat} -dT^{bat})\\
\frac{dS^{lat}}{dt}&=\frac{C^l}{T^{lat}}\frac{dT^{lat}}{dt}
\end{aligned}
\end{equation}
where $dS^{lat}$ is the change of entropy of the lattice, and $\kappa_b$ is the thermal link between the lattice and the bath. Together, Eq. \ref{4f_cross_relaxation}, \ref{4f_heat_equation} and \ref{lattice_heat_equation} describe completely the experimental setup. 

We note that all three sets of equations are {\it instantaneous}: thermodynamic relations in Eq. \ref{4f_cross_relaxation}--by its nature, and the heat flow equations Eq. \ref{4f_cross_relaxation}, \ref{4f_heat_equation}, by approximation that the heat conductance $\kappa_l$ and $\kappa_b$ are constants, exhibiting no frequency dispersion in the experimental frequency range. Therefore, we can use 
\begin{equation}\label{define_ac}
\begin{aligned}
    dH(t)&= dH(\omega) e^{-i\omega t}  \\
     dM(t)&= dM(\omega) e^{-i\omega t}  \\
   dS^{4f,lat}(t)&=dS^{4f,lat}(\omega)e^{-i\omega t}  \\
   dT^{4f,lat}(t)&=dT^{4f,lat}(\omega)e^{-i\omega t}  \\
\end{aligned} 
\end{equation}
to write a system of equations for frequency components, 
\begin{equation}\label{system_of_equations}
\begin{aligned}
   \left[ \begin{array}{l}
    dS(\omega) \\ dM(\omega) 
    \end{array} \right]^{4f} 
    = & 
 \left[   \begin{array}{cc}
     \frac{C}{T_0} & \gamma \\ \gamma & \chi 
    \end{array} \right]^{4f}
\left[     \begin{array}{l}
    dT(\omega) \\ dH(\omega) 
    \end{array} \right]^{4f}  \\
     \\
        -i\omega T^{4f}  dS^{4f}(\omega)   
    =& -\kappa_l (dT^{4f}(\omega)-dT^{lat}(\omega))  \\ 
-i\omega T^{lat} dS^{lat}(\omega) 
=&  \kappa_l (dT^{4f}(\omega)-dT^{lat}(\omega)) \\
&- \kappa_b (dT^{lat}(\omega)-dT^{bat}(\omega)). 
\end{aligned}
\end{equation}

Note that we assume the bath holds at constant temperature $dT^{bat}(\omega)=0$, and the temperature variation of each subsystem is much smaller than the total value (i.e. $T^{lat}\approx T_0$, $T^{4f}\approx T_0$, when one only considering the linear response of the setup.). Eq. \ref{system_of_equations} is a system of 4 equations for 4 unknowns ($dT^{4f}(\omega)$, $dT^{lat}(\omega)$, $dM(\omega)$ and $dS^{4f}(\omega)$). We will treat $dH(\omega)$ as a constant applied AC field, and $dS^{lat}(\omega)$ is completely defined by $(C^l/T^{lat})dT^{lat}(\omega)$.

We can now calculate two response functions, 
\begin{equation}
\begin{aligned}
\alpha(\omega) = \frac{ dM^{4f}(\omega)}{dH(\omega)}\,, \quad \text{and} \qquad 
\beta(\omega) = \frac{ dT^{lat}(\omega)}{dH(\omega)}    
\end{aligned}
\end{equation}
where $\alpha(\omega)$ is the effective magnetic response of the experimental setup, and $\beta(\omega)$ is the effective magnetocaloric response measured through the lattice subsystem. We note that due to the internal thermal exchange process between the pseudospin and lattice subsystems, $\beta(\omega)$ is no longer an intrinsic thermal response of the magnetic species. Solving for $\alpha(\omega)$ and $\beta(\omega)$ in Eq. \ref{system_of_equations}, we obtain:
 \begin{equation} \label{MCE_solution_1}
    \begin{aligned}
        &\alpha(\omega)=\\
        &\chi-\frac{T_0\gamma^2(i\omega)((\kappa_l+\kappa_b)-C^l(i\omega))}{-\kappa_l\kappa_b+(i\omega)(C^l\kappa_l+C^{4f}(\kappa_l+\kappa_b))-C^{4f}C^{l}(i\omega)^2} \\
    \end{aligned}
\end{equation}
 \begin{equation} \label{MCE_solution_2}
    \begin{aligned}
        &\beta(\omega) = \\
        &\frac{T_0 \gamma \kappa_l (i\omega)}{-\kappa_l \kappa_b+(C^{4f}+C^l)\kappa_l (i\omega)+C^{4f} \kappa_b(i\omega)-C^{4f} C^l (i\omega)^2} \\
    \end{aligned}
\end{equation}

The real and imaginary parts of $\alpha(\omega)$ and $\beta(\omega)$ are plotted in Fig. \ref{fig-1}(c, d). As we show below, these are quantitatively captured by measurements of \ce{YbVO4}. We note that the functional form of the $\beta(\omega)$, which is caused by the AC MCE effect, also depicts a portion of a circular feature in the Cole-Cole plot, similar to the appearance of the $\alpha(\omega)$ measurement. Looking at the solutions more carefully, the peak in the real part of the $\beta(\omega)$ is found to occur at a frequency that reflects the geometric mean of the thermal relaxation times $\tau_{4f}=C^{4f}/\kappa_l$ and $\tau_{ext}=C^{4f}/\kappa_b$ \footnote{In principle, we can also use  $C^{lat}$ to define the $\tau_{ext}$. We eventually chose to use $C^{4f}$ instead of $C^{lat}$ for simplicity during fitting analysis.}, reinforcing the points made in previous sections that external relaxation effects can complicate the intepretation of AC measurements.

\begin{figure}[ht!]
	\includegraphics[width = \columnwidth]{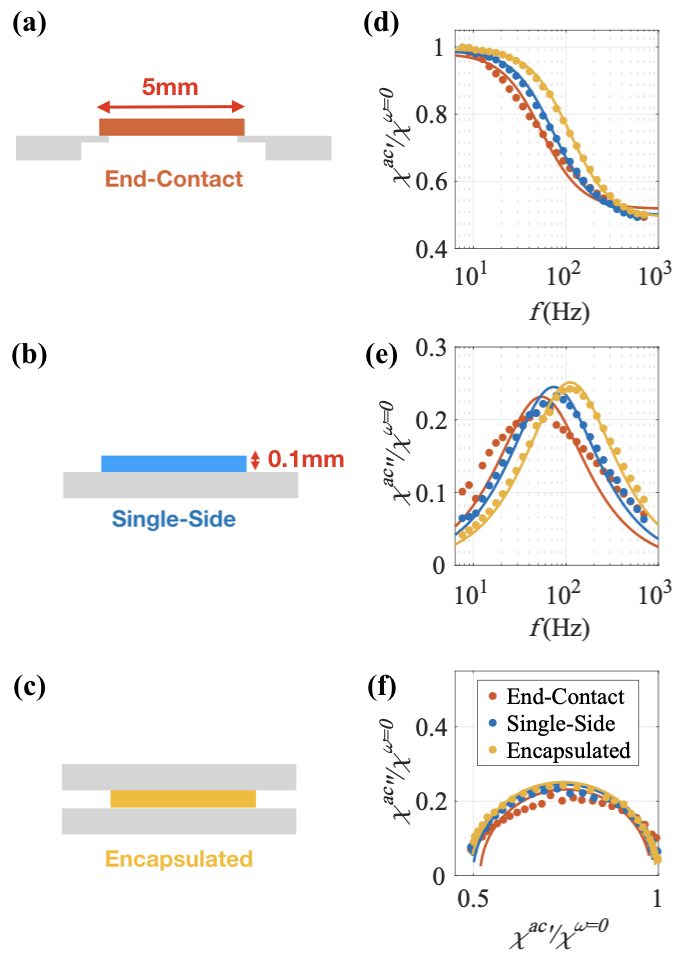}
	\caption{\label{AC_chi} Data illustrating how sample mounting configurations can affect AC Magnetic relaxation. Panel (a-c) illustrates three different sample mounting configurations, described in the main text. Colored bars represent the sample that is to be measured, together with its dimensions. The same sample of \ce{YbVO4} is used for all three configurations to enable direct comparison. Grey blocks indicate quartz platforms that serve as heat baths. The sample is oriented with the magnetic field aligned along the long c-axis of the crystal (horizontal in the schematic diagrams). AC susceptibility measurements were made at 3 K, 0.1 T, using an AC field of 3 Oe. Panels (d) and (e) show the real ($\chi^{ac'}$) and imaginary parts ($\chi^{ac''}$) normalized by the fitting parameter ($\chi^{\omega=0}$) taken as the isothermal value ($\chi$ in Eq. \ref{Debye}). Panel (f) shows the associated Cole-Cole plot in which frequency is an implicit variable. Data are shown for the three mounting configurations indicated in panel (a), using the same colors to differentiate the three configurations. Solid lines in (d,e,f) are fit results based on Eq. \ref{Debye}. The encapsulated configuration (yellow data points) yields data that are closest to the idealized Debye relaxation conditions. }

\end{figure}

\section{AC magnetic response of \NoCaseChange{\ce{YbVO4}}}
\label{AC_magnetic_response}
Heat exchange between the magnetic system and its environment is neither temporally instantaneous nor spatially homogeneous for realistic mounting conditions. Consequently, the characteristic time and amount of the heat exchange and associated magnetic relaxation are influenced by both the driven excitation field and the extrinsic mounting conditions, causing the time-dependent magnetic relaxation behavior not to be described by a uniform thermodynamic character. Moreover, when an internal process exists, the thermodynamic condition of each subsystem has to be described separately, making the overall quasi-adiabatic response unknown.


If the thermal response described above is neglected, then the measured $\chi^{ac}$ is often reported as the true AC susceptibility $\chi^{ac}$, though this is done incorrectly if one treats $\chi^{ac}$ as the intrinsic spin dynamics. Moreover, assuming the intrinsic spin susceptibility $\chi$ has no frequency dependence (i.e. $\chi^{ac}$ obtains all of its frequency dependence through interaction with other subsystems and/or the thermal bath), then the zero frequency value of $\alpha(\omega)=\chi$, and $\alpha(\omega)$ describes fully the frequency dependence of the magnetic response. Hence, we use an approximately equal sign in Eq. \ref{Debye} to associate the measured magnetic response to the commonly reported AC susceptibility $\chi^{ac}$.

The above insight indicates that the actual thermal response of a material cannot, and indeed must not, be neglected in the measurement of the magnetic response at finite frequency. This motivates a more careful investigation of the cross-relaxation that occurs in the experiment.

First, we examine the consequence of Eq. \ref{MCE_solution_1}, which can also give a commonly observed magnetic response described by a Debye-relaxation behavior\cite{Topping_2019}. It is known that when a single relaxation time $\tau$ governs the periodic flow of energy in a system, the dynamical susceptibility as a function of driven frequency $\omega$ is given by a Debye-like relaxation process in various reported AC magnetic susceptibility measurements\cite{Topping_2019}. Here, we realize that such a Debye-like relaxation process can be described exactly by Eq. \ref{MCE_solution_1} after being simplified to the following form:

\begin{equation} \label{Debye}
    \begin{aligned}
    \alpha(\omega)\approx\chi^{ac}(\omega)&=\chi-\frac{T_0\gamma^2(-i\omega)}{C^{4f}(-i\omega)+\kappa_l}\\
    \end{aligned}
	\end{equation}

Eq. \ref{Debye} is the limit of Eq. \ref{MCE_solution_1}, when $\kappa_b$ goes to zero, and $C^l$ much bigger than $C^{4f}$ (see Appendix.\ref{app:Simplified-Thermal} for detailed reasoning of applying the approximation in real experimental setup).

We then examine the relation between the measured $\alpha(\omega)$ of a single crystal of \ce{YbVO4}, with the aim of demonstrating an important point, namely that the measured value can depend on the experimental configurations that are used, even when internal relaxation processes dominate the quasi-adiabatic response.

To illustrate the real effects of extrinsic factors that indirectly affect the $\alpha(\omega)$ of \ce{YbVO4}, we first compare the magnetic relaxation behavior in different mounting configurations in Fig. \ref{AC_chi}, where the same sample mounted in different conditions are listed in Fig. \ref{AC_chi}(a). In the first case of end-contact (blue), the sample's extrinsic thermal contact is minimized on both ends using GE Low-Temperature Varnish (GE Varnish) and Teflon materials with relatively low thermal conductivity at low temperatures. In the second case of the single-side mounting condition (red), the sample has one side connected to the sample holder via GE Varnish. This is consistent with the standard procedure recommended by the MPMS system vendor. In the third case, the extrinsic thermal contact is maximized by encapsulating the sample between two quartz surfaces (yellow). Fig. \ref{AC_chi}(b,c) plots the magnetic relaxation behaviors from $\alpha(\omega)$. 

A comparison of the measured susceptibility for the three different extrinsic mounting conditions reveals that the adiabatic (high frequency limit) and isothermal (low frequency limit) susceptibilities are independent of the extrinsic condition. Similarly, the magnitude of the excitation field does not affect the adiabatic and isothermal response (we varied the amplitude of the driven field between 1 Oe and 10 Oe, finding that this did not affect the result). This is in contrast with expectations if the relaxation is dominated by external relaxation. 


To further visualize the relaxation behavior, the real part of the $\alpha(\omega)$ is plotted against its imaginary part (known as Cole-Cole plot) in Fig. \ref{AC_chi}(f). Here, we observe that the measurement under good thermal conditions gives an undistorted semicircle with an aspect ratio of 1, while the other two cases slightly deviate from an undistorted circle. The result implies that optimized thermal condition with a uniform thermal contact enables us to further conduct quantitative analysis on the magnetic relaxation behavior of the \ce{YbVO4}. Conversely, poor thermal contact, which results in magnetization and temperature oscillations that are more spatially and temporally inhomogeneous, leads to a more complex signature that does not easily reflect the actual quasi-adiabatic response of the material.

A fitting of the Debye-like relaxation behavior based on Eq. \ref{Debye} was performed for the three different mounting conditions in Fig.\ref{AC_chi} gives the best fit to experimental data in the encapsulated mounting condition in Fig. \ref{AC_chi}, when the sample is optimally thermalized to the bath (yellow data points).  

\section{AC magnetocaloric response of \NoCaseChange{\ce{YbVO4}}}

\begin{figure}[tp]
	\includegraphics[width = \columnwidth]{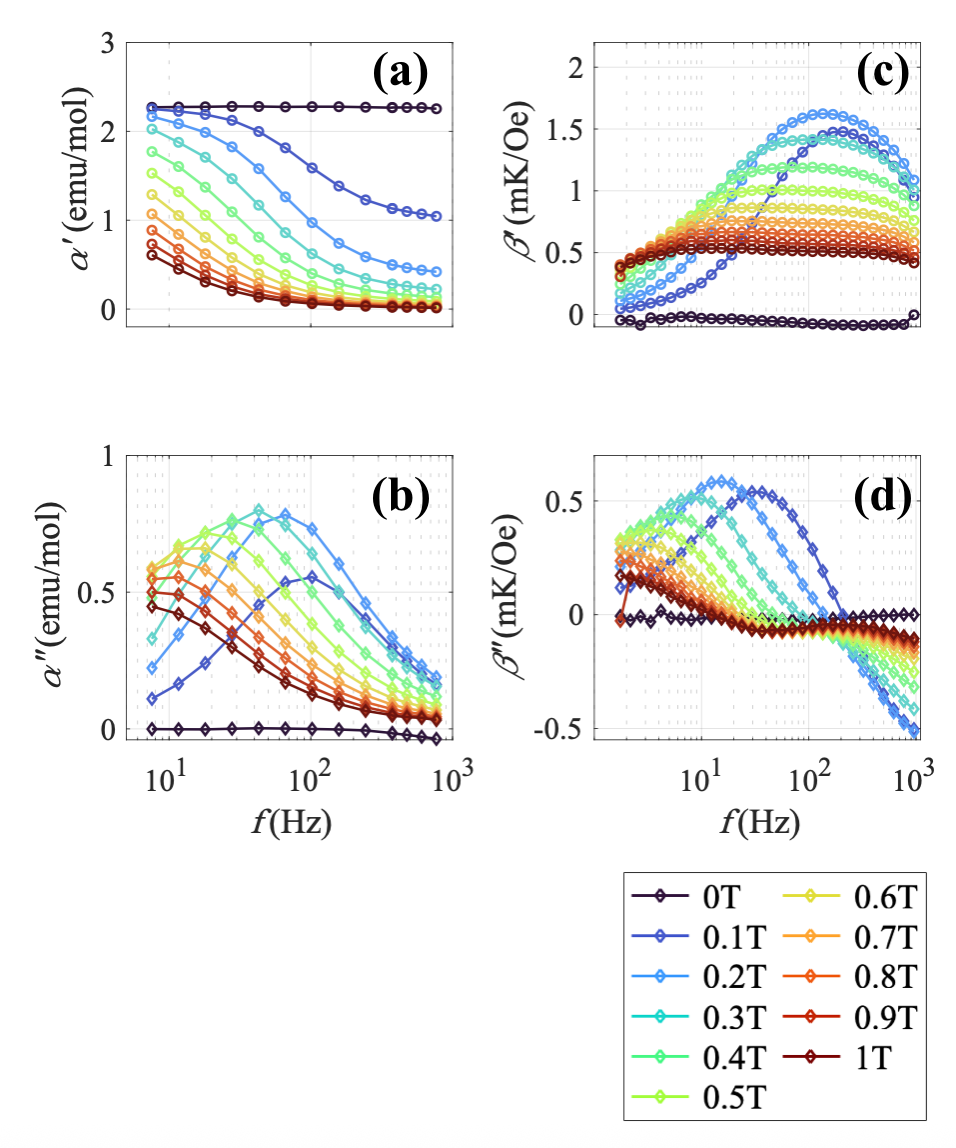}  
	\caption{\label{YbVO4_exps} Experimental results for \ce{YbVO4} showing the frequency dependence of the real ($\alpha'$) and imaginary ($\alpha''$) parts of (a,b) $
    \alpha$ and (c,d) the AC MCE ($\beta$) at 3 K. Data are shown for representative DC magnetic fields from 0 T to 1 T.}
\end{figure}

Fig. \ref{YbVO4_exps} shows the measured frequency-dependence of $\alpha(\omega)$ and the AC MCE of \ce{YbVO4} at 3 K for magnetic fields between 0 and 1 T. An AC field of 3 Oe was used. In order to ensure meaningful comparison of the two measurements, the same sample, and the same mounting configuration (corresponding to the yellow configuration shown in Fig. \ref{AC_chi}), were used for both measurements.  

Before performing detailed fits to the data based on the thermal model, we first note three observations that are evident simply by inspecting the data shown in Fig. \ref{YbVO4_exps}. First, the data clearly follow the anticipated functional forms sketched in Fig. \ref{fig-1}(c, d). Second, there is clearly a strong field dependence to both quantities, as anticipated for \ce{YbVO4} due to Zeeman splitting of the CEF eigenstates. And third, the field-dependence is very similar for the two quantities; i.e. the maximum in the real part of $\beta$ follows the same trend as the maximum in the imaginary part of $\alpha$. The consistency in the shift of the characteristic frequency between the two measurements implies that the same internal and extrinsic heat transfer processes are being captured in both experiments.

We note that such highly frequency-dependent responses are not observed in \ce{GdVO4} (see Appendix. \ref{app:GdVO4_exp}), a material for which phonon bottleneck effects are not anticipated due to the absence of CEF effects, even when it is held under similar experimental conditions (i.e. similar external thermal relaxation). Thus, the relaxation effects evident in Fig. \ref{YbVO4_exps} point to an intrinsic slow relaxation, as anticipated for \ce{YbVO4}, with the anticipated strong field dependence.

\section{\NoCaseChange{Fits to the thermal model}}

We will be interested in two responses of this system $\alpha(\omega)$ and $\beta(\omega)$ that can be measured directly in an experimental setup.

\begin{figure}[t]
	\includegraphics[width = \columnwidth]{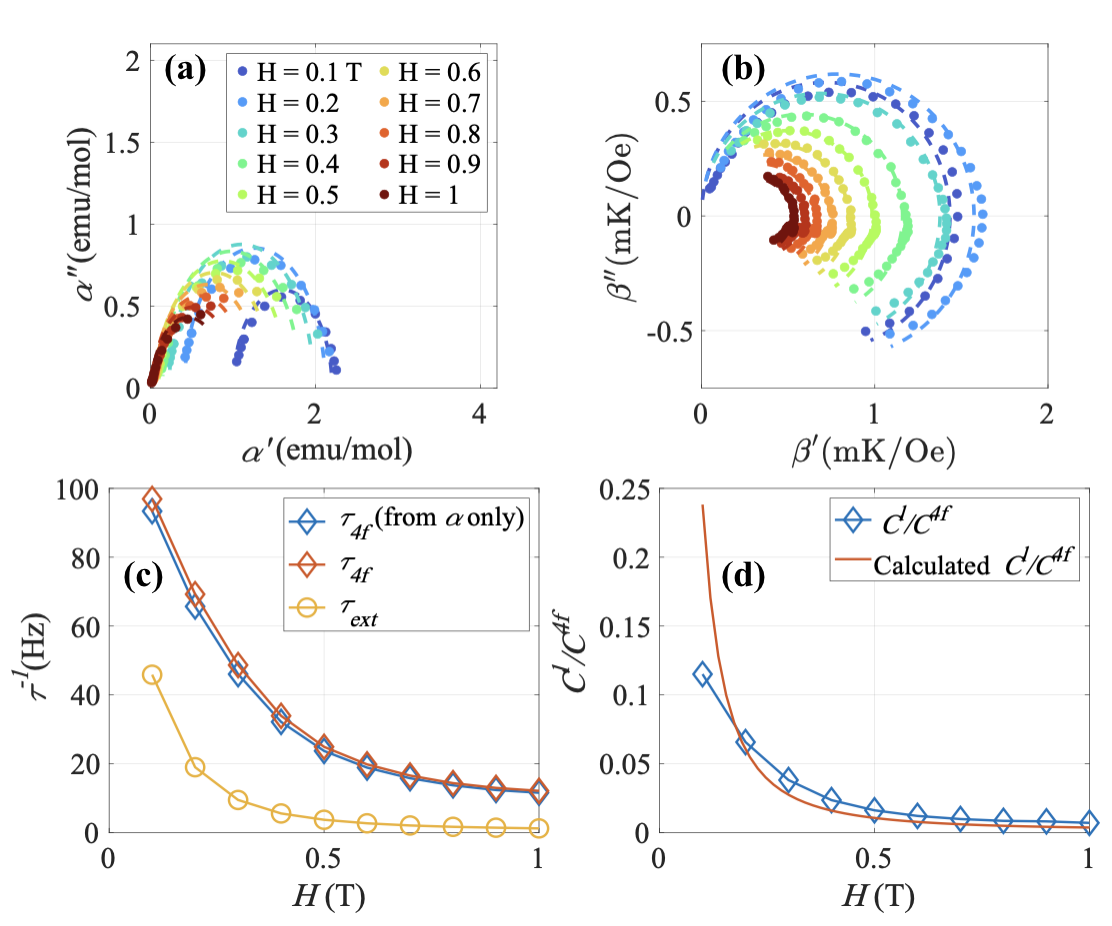}
	\caption{Fitting of \ce{YbVO4} data with Eq. \ref{fit_equation_chi} and \ref{fit_equation_MCE} measured at 3K. (a) A Cole-Cole plot showing the real ($\alpha$) against the imaginary part ($\alpha$) of the dynamical susceptibility. (b) A similar plot for the AC MCE, where $\beta$. (c) The characteristic spin-lattice relaxation time ($\tau_{4f}$ and $\tau_{ext}$) obtained from the fits, as described in the main text. (d) The parameter $C^l/C^{4f}$ from the fit result, compared with the calculated value based on the Schottky anomaly of the ground-state Kramers doublet in \ce{YbVO4} and the phonon background described by the Debye Model, as described in the main text. }\label{fitresult}
\end{figure}

The data shown in Fig. \ref{YbVO4_exps} can be fit to the solution provided by Eq. \ref{Debye} and \ref{MCE_solution_2} to reveal the field dependence the two time constants. (i.e. the intrinsic $\tau_{4f}=C^{4f}/\kappa_l$ and the extrinsic thermal relaxation $\tau_{ext}=C^{4f}/\kappa_b$). The fits, which we describe in greater detail below, are shown in Fig. \ref{fitresult} (a) and (b) as dashed lines on the associated Cole-Cole plots.

To simplify the fitting procedure, four fit parameters can be defined: $B=(T\gamma^2)\kappa_l$, $\tau_{ext}=C^{4f}/\kappa_b$, $\tau_{4f}=C^{4f}/\kappa_l$, and $\eta=C^l/C^{4f}$, where $\gamma$ is the magnetocaloric coefficient defined in Eq \ref{4f_cross_relaxation}. From experimental observations and fit results, we find that the effect of the thermometer does not significantly affect the in-phase component $\beta'(\omega)$, but does contribute a small prefactor to the out-of-phase component $\beta''(\omega)$, resulting in a negative value for the fit parameter $A$ with magnitude slightly less than 1. A more detailed discussion of the effect of the thermal sensor, together with experimental characterization, is given in the Appendix. \ref{app:Thermometer}. Rewriting the four fitting equations after substitution of the redefined fitting parameters to Eq \ref{Debye} and \ref{MCE_solution_2} yields:

\begin{equation} \label{fit_equation_chi}
    \begin{aligned}
        \alpha'(\omega)&=\chi-\frac{B \omega ^2}{\frac{1}{\tau_{4f}}+\tau_{4f}  \omega ^2}\\
         \alpha''(\omega)&=\frac{B \omega }{1+\tau_{4f}^2\omega ^2}\\
    \end{aligned}
	\end{equation}

\begin{equation} \label{fit_equation_MCE}
    \begin{aligned}
         &\beta'(\omega)=\\
         &\frac{\gamma B\tau_{ext}\omega(1-\eta \tau_{4f} \tau_{ext} \omega^2)}{\tau_{4f}+\tau_{4f}(\tau_{4f}^2+2\tau_{4f}\tau_{ext}+(1+\eta)^2\tau_{ext}^2)\omega^2+\eta^2\tau_{4f}^3\tau_{ext}^2\omega^4}\\
         &\beta''(\omega)=\\
         &\frac{\gamma AB\tau_{ext}(\tau_{4f}+\tau_{ext}(1+\eta))\omega^2}{\tau_{4f}+\tau_{4f}(\tau^2+2\tau_{4f}\tau_{ext}+(1+\eta)^2\tau_{ext}^2)\omega^2+\eta^2\tau_{4f}^3\tau_{ext}^2\omega^4}\\
    \end{aligned}
\end{equation}

We should note that $\alpha(\omega)$ does not describe the intrinsic $(\chi)$ of the spin subsystem in Eq. \ref {4f_cross_relaxation}. Rather, it describes an effective response specific to the energy exchange. In particular, as will be shown later, the magnetic susceptibility of the 4f spin is static, i.e., it does not exhibit dispersion within the frequency range measured.  

 One final aspect of the fitting procedure accounts for the presence of the thermal sensor in the MCE measurement. The observed aspect ratio of the Cole-Cole plot of the AC MCE experiment is very slightly distorted from being a perfect complete circle predicted by Eq. \ref{MCE_solution_2}. We attribute the deviation to the effect of the thermometer that is attached to the sample, which slightly changes the thermal configuration compared to that of the $\chi^{ac}$ experiment, where a sensor is not thermally attached to the sample.

Returning to the fit results shown in Fig. \ref{fitresult} panels (a) and (b), dashed lines indicate the fits of measured $\alpha$ and $\beta$ based on Eq. \ref{fit_equation_chi} and \ref{fit_equation_MCE}. The field dependence of the two time constants that are extracted from these fits is shown in panel (c). The red data points show the results for $\tau_{4f}$ when both $\alpha$ and $\beta$ are simultaneously fit by the full thermal model involving two time constants. Yellow data points show the associated $\tau_{ext}$ values from those same fits to the full thermal model. In order to confirm that the combination fitting method is valid, we also fit the measured $\alpha$ separately using only Eq. \ref{fit_equation_chi}. Blue data points show the result from fitting $\alpha$ using only a single-time constant $\tau_{4f}$. The goodness of fits in panels (a) and (b), combined with the internal consistency of the two fitting methods shown in panel (c), imply that the frequency characteristics of the two experiments capture the same energy transfer processes very well.

 To further confirm that the fit result accurately describes an internal energy exchange process between the spin and lattice components, we compare the fitted value of heat capacity ratio $\eta=\frac{C^{l}}{C^{4f}}$ with the calculated value for a 2-level spin system and a phonon bath with $T^3$ dependence. The fit parameter that describes the heat capacity ratios is shown in Fig. \ref{fitresult}(d), and it is indeed consistent with the calculated value.

\section{\NoCaseChange{Discussion}}
It is instructive to consider the relation between the thermal analysis that we present above to a generalized linear response theory\cite{Landau, Onsager}. When multiple variables interact with each other in a circuit, the overall relaxation behavior is characterized by a set of thermodynamic conjugate variables, which refer to pairs of ``force ($f_\omega$)" and ``displacement($x_i(\omega)$)" that respond directly to each other. A change in one variable directly affects the other. The application of ``forces" results in a corresponding ``displacement" that characterizes the linear response function $\alpha_i(\omega)$:

\begin{equation} \label{conjugate_responce}
    \begin{aligned}
        x_i(\omega)=\alpha_i(\omega) f_\omega
    \end{aligned}
	\end{equation}

The most commonly existing pair of conjugate variables is temperature and entropy $(T, S)$. When the thermodynamic state of a closed system remains unchanged by applied fields, the thermal-susceptibility is uniquely defined by the heat capacity because $ dS=(C_p/T) dT$. Consequently, the dynamical response is completely determined by the rate of thermal conductance, affected by an external heat source. If, however, the system develops a caloric response, it must imply that another pair of conjugate variables exists in at least one of the circuit components. For example, the measurement results above contain 3 pairs of conjugate variables, which are $(H, M)$, $(T^{4f}, S^{4f})$ and $(T^{lat}, S^{lat})$. Consequently, the full relaxation behavior can be completely determined by a linear response matrix:

\[
\begin{blockarray}{cccc}
& M & S^{4f} & S^{lat} \\
\begin{block}{c(ccc)}
  H & \alpha_{11} & \alpha_{12} & \alpha_{13} \\
  T^{4f} & \alpha_{21} & \alpha_{22} & \alpha_{23} \\
  T^{lat\,} & \alpha_{31} & \alpha_{32} & \alpha_{33} \\
\end{block}
\end{blockarray}
 \]

The diagonal terms represent the susceptibility response. $\alpha_{11}$ is the AC magnetic susceptibility ($\alpha(\omega)$ in Eq. \ref{MCE_solution_1}), $\alpha_{22}$ accounts for the thermal susceptibility of the spin affected by the magnetic field, and $\alpha_{33}$ accounts for thermal susceptibility of the lattice. From an experimental perspective, we note that any conventional dynamical measurement only obtains the frequency response function of the diagonal matrix element. Although all matrix elements must satisfy the relation of $\alpha_{i,i+j}\alpha_{i+j,i}=\alpha_{i,i}\alpha_{i+j,i+j}$, the matrix cannot be uniquely defined by dynamical susceptibility only. Because the term $\alpha_{33}$ is special in the sense that it is a single positive real constant $C^{l}/T_0$ instead of a complex response function. Therefore, the AC caloric effect function $\beta(\omega)$ can be associatd with $\alpha_{13}$ by the relation $\beta(\omega)=\alpha_{13}/\alpha_{33}$. 

The matrix response shown above provides an illustrative example that any dynamical behavior of a non-instantaneous magnetic response is most likely to be the complex result of both internal structure and external conditions. Hence, for systems that are composed of multiple degrees of freedom, the measured $\chi^{ac}$ on its own is insufficient to reflect a complete dynamical process. The discretized thermal model based on spin-lattice relaxation underscores the importance of measuring the AC caloric response as a reciprocal observation to provide complementary information to any susceptibility analysis.
 
\section{\NoCaseChange{Summary}}

We have developed an experimental method to measure the AC MCE, and have introduced a discretized thermal analog circuit approach that fully describes the cross-relaxation between magnetization and temperature in the frequency domain in the presence of an AC magnetic field excitation. This approach permits a full understanding of the frequency dependence of $\chi^{ac}$ and the AC MCE through the presence of both extrinsic and intrinsic relaxation processes
. 

We demonstrated the technique and the associated fitting methods using \ce{YbVO4}, a material for which there are slow magnetic dynamics at low temperatures arising from a phonon-bottleneck effect.

The magnetic dynamics of a wide variety of materials are often inferred from just susceptibility measurements, and analysed using a Debye, or closely related, model. Associated slow dynamics can be attributed to various physical effects. While these analyses are well-motivated, they are necessarily incomplete if cross-relaxation is neglected, potentially missing or mis-characterizing important new material-specific insights. The new caloric approach developed in this paper could provide additional evidence about these internal relaxation phenomena, while also providing a grounded description of the accompanying thermal relaxation effects associated with extrinsic effects. We hope that the approach we have outlined here will prove helpful in future studies of magnetic dynamics in a wide range of materials.

\section{\NoCaseChange{Acknowledgements}}
We thank Matthias S. Ikeda for fruitful discussions and original insights on the AC caloric measurement techniques. Low-temperature measurements performed at Stanford University were supported by the Air Force Office of Scientific Research Award FA9550-24-1-0357, using a cryostat acquired with award FA9550-22-1-0084. Crystal-growth experiments were supported by the Gordon and Betty Moore Foundation Emergent Phenomena in Quantum Systems Initiative Grant GBMF9068. Work at Los Alamos National Laboratory is supported by the NSF through DMR-1644779 and DMR-2128556, the U.S. Department of Energy, and the DOE/BES “Science of 100 T” grant. M.P.Z. was also partially supported by a National Science Foundation Graduate Research Fellowship under grant number DGE-1656518. L.Y. acknowledges support from the Marvin Chodorow Postdoctoral Fellowship at the Department of Applied Physics, Stanford University. 

\begin{appendices}

\section{\NoCaseChange{Experimental set up of the AC MCE measurement} \label{app:mce_exp}}

An electronic transport measurement of a resistance bridge with a dual-frequency lock-in technique is introduced to detect and measure a temperature oscillation on the order of milliKelvins. The setup consists of the driven magnetic coil, a Stanford Research SR860 lock-in amplifier in the dual mode (measures bridge voltage $V_b$), and another SR860 in the external mode (measures the sample's thermometer resistance $V_S$). The amplified driven coil signal gives an additional isolated output via the Electronic Module of the MPMS System. The voltage output with the internal reference frequency of the first lock-in was amplified into a current signal via a Stanford Research CS580 Voltage Controlled Current Source and also gave the reference signal input of the second lock-in. The Electron Transport Option (ETO) probe provided by the MPMS accessory kits was applied to enable transport measurement.

\begin{figure}[tp]
	\includegraphics[width = \columnwidth]{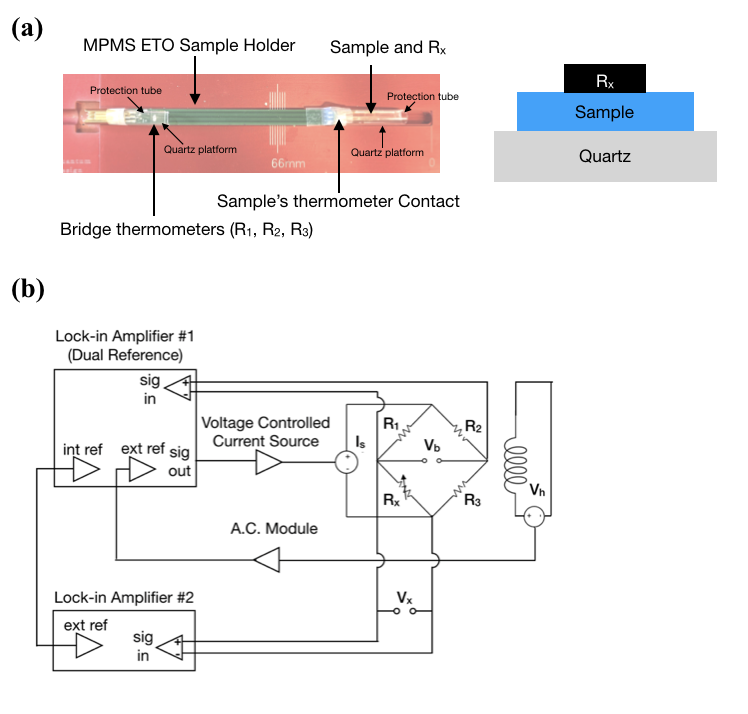}  
	\caption{\label{fig-7} (a) Actual photo of an AC MCE probe device assembly and a schematic showing the stacking sequence between thermometer $R_x$, sample, and the holder. The sample is located at the end quartz component that is attached to the ETO probe head. Before measurement, the center of the sample is aligned with the center of the applied magnetic field via an automatic sequence}. (b) Circuit diagram of the AC MCE measurement. Here, $R_x$ represents the thermal resistor. $R_1$, $R_2$, and $R_3$ are bridge resistors. The raw temperature oscillation signal is obtained from the bridge voltage $V_b$ via a lock-in amplifier in a dual mode with its internal reference, and external reference from the magnetic coil ($V_h$). A second lock-in amplifier measures $V_x$ to obtain the temperature profile of the thermal resistors. A Voltage Controlled Current Source converts the $1\ \text{V}$ voltage output of the lock-in amplifier to an excitation current of $100 \ \mu\text{A}$ to the Wheatstone Resistor Bridge.
 
\end{figure}

\section{\NoCaseChange{Energy exchange of transitions between CEF states in \ce{YbVO4}} \label{app:CEF}}

In order to identify the transition between the magnetic induced splitting between the ground-state doublet in \ce{YbVO4}, we consider the electron of the 4f shell embedded in a rare-earth vanadates crystal structure with $D_{2d}$ symmetry. The corresponding CEF Hamiltonian can be written as:
\begin{equation} \label{CEF_Hamiltonian}
\begin{aligned}
	H_{CEF}=B_2^0 O_2^0+B_4^0 O_4^0+B_4^4 O_4^4+B_6^0 O_6^0+B_6^4 O_6^4\\
 \end{aligned}
\end{equation}
where $B_n^m$ are the CEF parameters defined by the \ce{RVO4} lattice, and $O_n^m$ are Steven operators. The irreducible representations of the ground-state, first and second excited states calculated for the \ce{Yb^{3+}} ion with crystal field parameter \cite{NIPKO1997} are:

\begin{equation} \label{CEF_Eignestates}
\begin{aligned}
	\pm\Gamma_1^7&=0.901\ket{\pm\frac{7}{2}}-0.433\ket{\mp\frac{1}{2}}, E=0 \: meV\\
 \pm\Gamma_1^6&=\pm0.886\ket{\pm\frac{3}{2}}\mp0.464\ket{\mp\frac{5}{2}}, E=7.2 \: meV\\
 \pm\Gamma_2^7&=0.433\ket{\pm\frac{7}{2}}+0.901\ket{\mp\frac{1}{2}}, E=34.8 \: meV\\
 \end{aligned}
\end{equation}

By calculating the transitions with $\Braket{\Gamma| H_{CEF}|\Gamma}$ between the ground states and the two excited states, we found that the only transition between ($+\Gamma_1^7,+\Gamma_2^7$) and ($-\Gamma_1^7,-\Gamma_2^7$) are allowed, indicted by a non-zero value. Aside from crystal symmetry considerations, phonon modes investigated by room temperature Raman spectra \cite{Santos2007, SANTOS2007_2} and inelastic neutron scattering \cite{NIPKO1997} that break the $D_{2d}$ symmetry are being reported. Therefore, we have also taken consideration of the extra Stephen operators $\Braket{\Gamma| O_m^n| \Gamma}$ for $m=2,4,6$ and $n=0,2,4,6$ between each of the 6 total states above, we found that the extended CEF parameters allow transition between ($+\Gamma_1^7,-\Gamma_1^6$) and ($-\Gamma_1^7,+\Gamma_1^6$). There is no allowed transition between the ground state doublet $\Braket{\Gamma_1^7| O_m^n| -\Gamma_1^7}=0$.

\section{\NoCaseChange{AC magnetic susceptibility and AC MCE of \ce{GdVO4}}}
\label{app:GdVO4_exp}
\begin{figure}[t!]
	\includegraphics[width = \columnwidth]{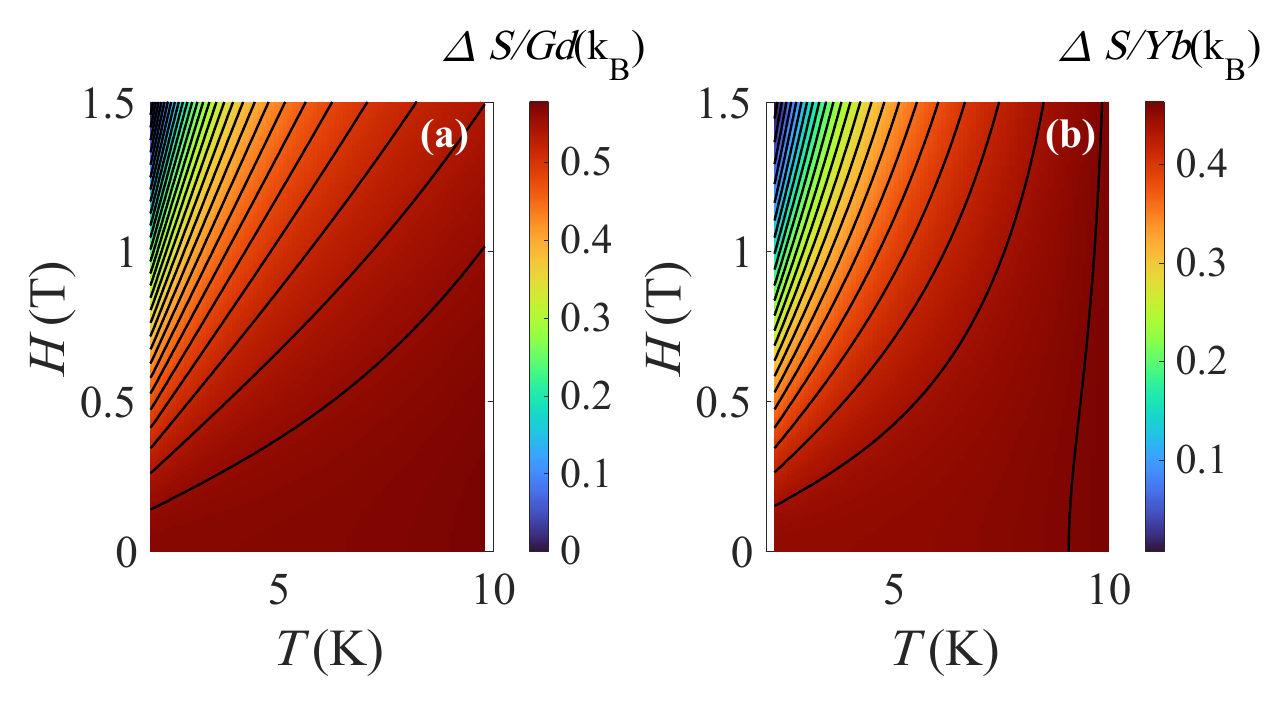}
	\caption{\label{entropy} The calculated entropy change of (a) \ce{GdVO4} and (b) \ce{YbVO4} with respect to the entropy value of 2 K and 0 T considered the total entropy of the lattice and the $4f$ contribution. The black lines represent isentropic contours.}
\end{figure}

Both \ce{YbVO4} and \ce{GdVO4} exhibit a large magnetocaloric effect at low temperatures. Compared with \ce{YbVO4}, no magnetic relaxation effect is observed in \ce{GdVO4} at low temperatures. Fig. \ref{entropy} illustrates the calculated entropy landscapes for both compounds. These landscapes are characterized by a phonon background and rare-earth degree of freedom, exhibiting comparable temperature changes when referenced at 2K and 0T. 

\begin{figure}[t!]
	\includegraphics[width = \columnwidth]{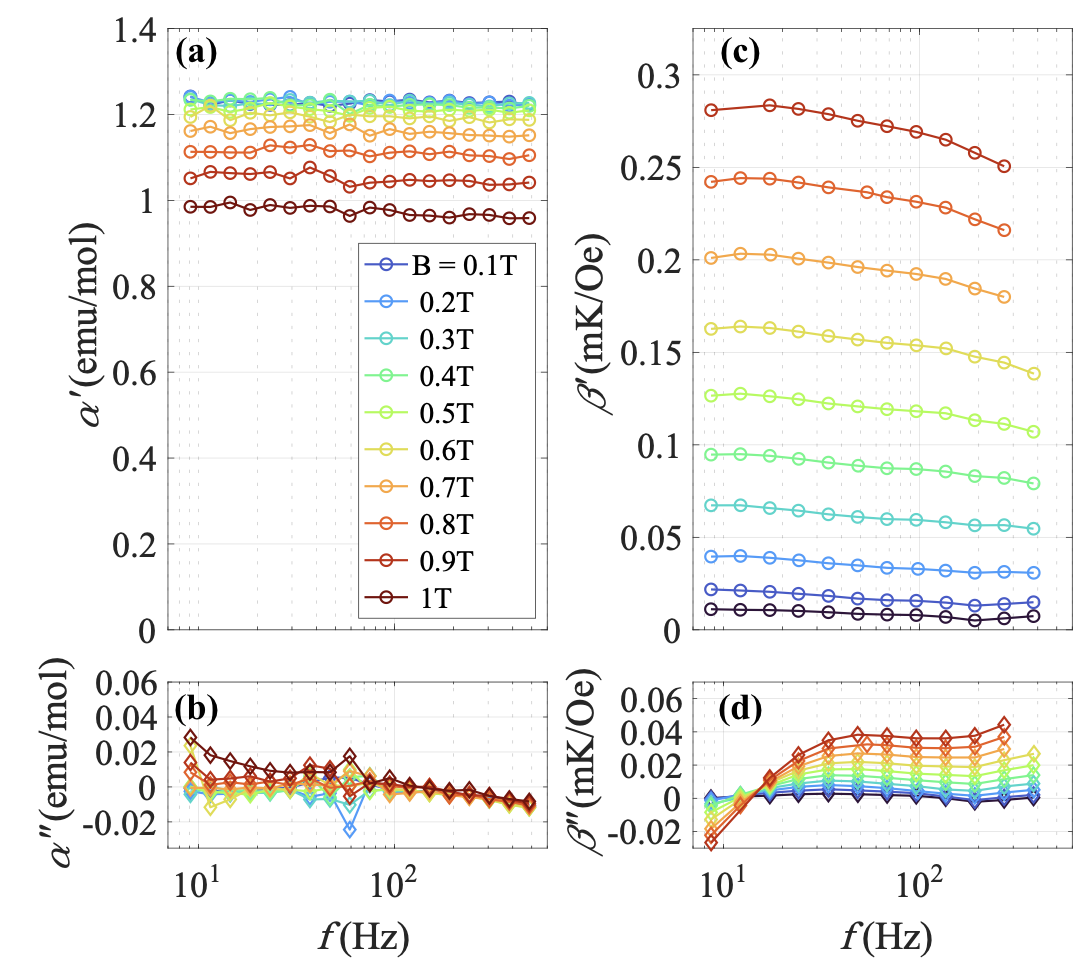}  
	\caption{\label{GdVO4_exps} (a,b) Real and imaginary part of the AC susceptibility of \ce{GdVO4}. Frequency-dependent results are plotted under different DC magnetic fields from 0 T to 0.18 T at 3 K. (c,d) DC magnetic field dependence of the AC MCE effect under the same field, frequency, and AC amplitude as (a,b).}
\end{figure}

\ce{Gd^{3+}} ion in \ce{GdVO4} does not split and remains an octuplet at low temperature. \ce{GdVO4} undergoes an antiferromagnetic transition at 2.5K, followed by a Spin-flop phase down to 1.37 K \cite{Mangum1972}. Consequently, \ce{GdVO4} will experience much less effect of such indirect transition due to much less magnetic anisotropy and very close Zeeman splitting between the CEF states. In Fig. \ref{GdVO4_exps}, we show the measured $\alpha$ and $\beta$ at 3 K. The external magnetic field is composed of a time-independent field and an AC excitation field. Notably, no significant frequency dependence was observed within the frequency range below 1 T, and the amplitude of the out-of-phase component remained comparable to the background noise for frequencies below 100 Hz during the measurements. The result implies an example that extrinsic time constants from both measurements exceed the maximum accessible frequency of the AC experiments, as an illustrative example of a material in the absence of internal dynamics.

\section{\NoCaseChange{Comparison with thermal conductivity measurements} \label{app:Extrinsic_Kappa}}
The fitting of $\alpha$ in the main text assumes that the lattice is well equilibrated with the sample holder, resulting in uniform temperature spatial distributions and a time constant $\tau_{ext}$ much smaller than the spin-lattice relaxation rate. This section confirms that the sample geometry and heat diffusion constant of rare-earth vanadates support the assumption with experimental data.

For the reason of limited data availability, we take an example of thermal properties of \ce{TmVO4} to estimate the characteristic time associated with the thermal diffusion constant in the same sample geometry as the "single-side" case in Fig. \ref{fig-1} (a). The thickness of the sample $l$ is 0.1 mm. The numeric values calculated in this section are all rough estimations and cannot be applied as a reference for accurate calculation. According to Ref. \cite{tmvo42022} and Ref. \cite{Vallipuram2024}, the thermal conductivity of \ce{TmVO4} is close to the value of $\kappa=0.35 \ \text{W/K\textsuperscript{2} m}$, and keeps almost unchanged up to the higher field. and the heat capacity is close to the value of $0.4  \ \textnormal{J/mol K}$ at 3 K and 0 T. Such value will be increased to roughly $2.9  \ \textnormal{J/mol K}$ at 3 K and 1 T. By calculating the diffusion constant $D=\kappa/C_p^{Vol}$, and the characteristic frequency $f=D/l^2$, the extrinsic time constant is $1.3 \ 10^4   \ \textnormal{Hz}$ at a field close to zero, and $1.7 \ 10^3  \ \textnormal{Hz}$ at 1 T. This value is a few orders of magnitude larger than the internal frequency observed in the magnetic relaxation measurements.

\section{\NoCaseChange{Simplified Thermal Exchange Model} \label{app:Simplified-Thermal}}

In this section, we explain the statement in Section.\ref{AC_magnetic_response}, which gives the simplified expression of $\alpha(\omega)$ for a real experimental setup. We consider a scenario when the lattice bath and the heat bath in Fig. \ref{fig-1}(b) have good thermal contact, hence, the two components are considered as a single environment relative to the 4f spin subsystem. By assuming the magnetocaloric effect heats the sample uniformly in ideal thermal conditions of a 2-components heat exchange model (i.e., neglecting the kind of extrinsic mounting effects described in the previous section), the heat transfer between the spin subsystem and a constant bath is described by a thermal transfer equation:
\begin{equation} \label{simplified_thermal_model}
    \begin{aligned}
         \left[ \begin{array}{l}
    dS(\omega) \\ dM(\omega) 
    \end{array} \right]^{4f} 
    = & 
 \left[   \begin{array}{cc}
     \frac{C}{T} & \gamma \\ \gamma & \chi 
    \end{array} \right]^{4f}
\left[     \begin{array}{l}
    dT(\omega) \\ dH(\omega) 
    \end{array} \right]^{4f}  \\
     \\
        -i\omega T^{4f}  dS^{4f}(\omega)   
    =& -\kappa_l (dT^{4f}(\omega)-dT^{bat})  \\
    \end{aligned}
	\end{equation}
    
Holding $dT^{bat}=0$ and solving for Eq. \ref{simplified_thermal_model}, we have obtained a solution that is completely equavlent by Eq. \ref{Debye}:

\begin{equation} \label{simplified_thermal_model_solution}
    \begin{aligned}
    \alpha(\omega) = \frac{ dM^{4f}(\omega)}{dH(\omega)}=\chi-\frac{T_0\gamma^2(-i\omega)}{C^{4f}(-i\omega)+\kappa_l}\\
    \end{aligned}
	\end{equation}

The solution here is the same as the solution of $\alpha(\omega)$ solved from Eq. \ref{system_of_equations}, where we assume that the effective $\kappa_b$ goes to zero and the effective $C^l$ is much greater than $C^{4f}$ (which leads to Eq. \ref{Debye}). Although good extrinsic thermal contact in Eq. \ref{system_of_equations} implies a large value of $\kappa_b$, part of the sample mounting will be considered as a part of the $C^l$ as well in real experiment, consequently, the value of the effective $C^l$ in the heat exchange model will be larger than the expected value of the real $C^l$, and the effective $\kappa_b$ describes another extrinsic thermal conductivity attributed to the extended lattice subsystem. Therefore, the approximation of $C^l$, $\kappa_b$ in the main text and the simplified thermal exchange model are equivalent in explaining the Debye-like relaxation behavior observed in $\chi^{ac}$. Such ideas of an extended lattice bath cannot be considered in the AC MCE, because the thermometer measures a local region of the sample.


\begin{figure}[tp]
	\includegraphics[width = \columnwidth]{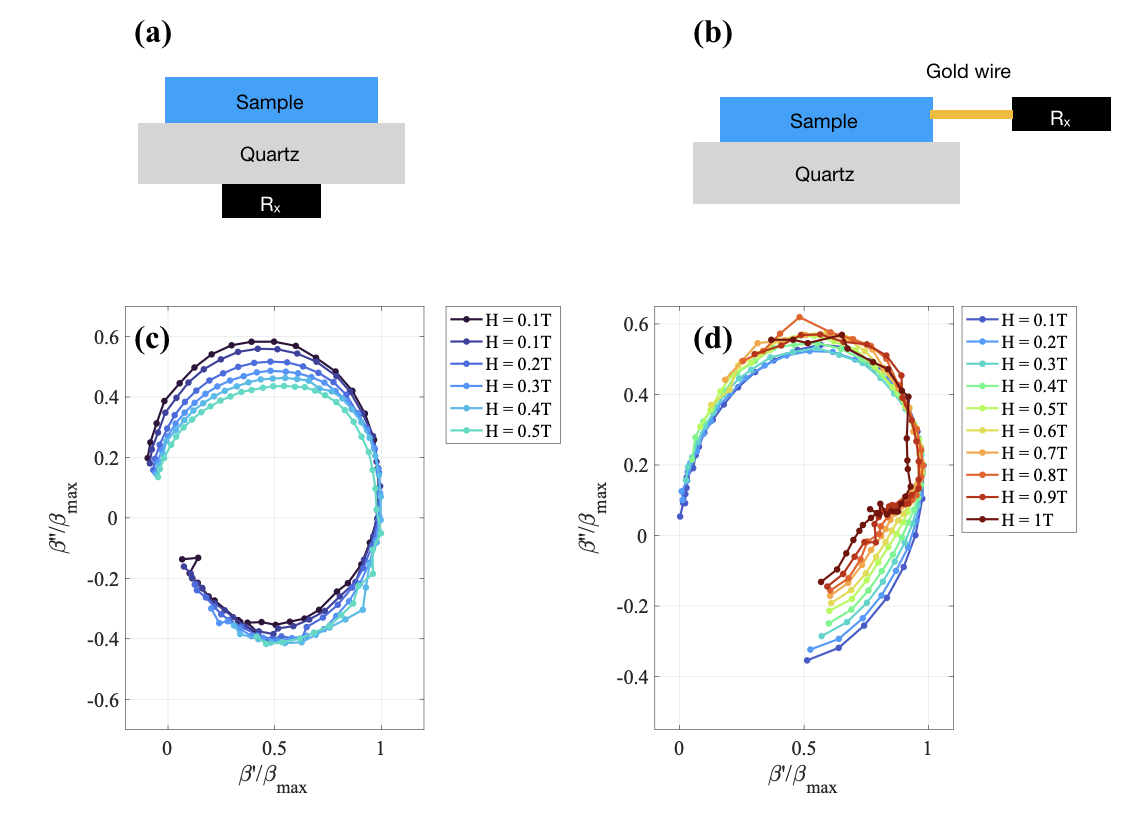}  
	\caption{\label{extrinsic_mce} (a,b) Schematics showing two additional sample mounting methods of AC MCE setup, as examples of sub-optimal mounting conditions which limited the thermal conduction rate between samples and thermometers. In (a), the thermometer and sample are separated by the quartz component, and in (b), they are separated by a 0.05 mm diameter gold wire served as a thermal lead. (c) The Cole-Cole plot of the measurement result $\beta/\beta_{max}$ of the mounted method in (a), where real and imaginary part of $\beta$ is normalized with a maximal value $\beta_{max}$ of each curve. (d) The Cole-Cole plot of the measurement result of the mounted method in (b).}
 
\end{figure}

\section{\NoCaseChange{Effect of thermometer and mounting condition in the AC MCE experiment} \label{app:Thermometer}}

The heat exchange model in the main text overlooks the effect of the thermometer becasue it has a marginal impact on the thermal exchange between several major components. To develop a generalized description of the thermometer reading that gives a conjugate value to the temperature response, we consider the simplest scenario:  A thermometer is attached to a part of a sample with thermal conductivity $\kappa_{\theta}$. If the sample and thermometer have a good contact, the third time constant $\tau_{\theta}$ described by $C^l/\kappa_{\theta}$ will be much smaller than $\tau_{4f}$ and $\tau_{ext}$. Hence, we consider the heat flow between a local part of the sample and the thermometer is described by a quasi-adiabatic condition where heat is exchanged between the sample and the thermometer only. Under the same fully discretized description that assumes the sample heats uniformly with a thermometer, the temperature of the sample and thermometer are defined as $T^{s}(\omega)$ and $T^{\theta}(\omega)$. According to the quasi-adiabatic condition, the only heat exchange equation can be written as:

\begin{equation} \label{eqn_heat_exchange}
    \begin{aligned}
    q_{\theta \leftarrow sample}&= -q_{sample \leftarrow \theta}\\
    &=-T^{\theta}\frac{dS^{\theta}(t)}{dt}\\
    &=-\kappa_{\theta}(dT^{\theta}(t)-dT^{sample}(t))\\
    \end{aligned}
	\end{equation}
    
when the steady-state change of entropy is periodic in both components, we can also write the equation in its frequency components:

\begin{equation} \label{eqn_heat_exchange_2}
    \begin{aligned}
   -i\omega dS^{\theta}(\omega)T^{\theta}&= i\omega dS^{s}(\omega)T^{s}\\
   &=-\kappa_{\theta}(dT^{\theta}(\omega)-dT^{sample}(\omega))\\
   dS^{sample}(\omega)&=\frac{C^{sample}}{T_0}dT^s(\omega)\\
   dS^{\theta}(\omega)&=\frac{C^{\theta}}{T_0}dT^{\theta}(\omega)
    \end{aligned}
	\end{equation}

The solution can be written in a matrix form as:
\[
\begin{blockarray}{ccc}
 & S^{s} & S^{\theta} \\
\begin{block}{c(cc)}
  T^{s \,} & \frac{C^{s}}{T_0} & -\frac{C^{\theta}}{T_0}\frac{\kappa_{\theta}}{\kappa_{\theta}-iC^{\theta}\omega}  \\
  T^{\theta \,} & -\frac{C^{s}}{T_0}\frac{2\kappa_{\theta}}{\kappa_{\theta}+iC^{\theta}\omega}  & \frac{C^{\theta}}{T_0} \\
\end{block}
\end{blockarray}
 \]
It is evident that the two off-diagonal terms have factors that are complex conjugate to each other, with their product equal to that of the two diagonal terms. This is because the complex relaxation functions have exactly the same time constant, ensuring that their product matches the product of the diagonal term, in accordance with the defining properties of the linear response matrix. It implies that when a circuit representing a thermometer is connected to a measured species, the sign of the imaginary part of the thermometer is inverted due to a conjugate relationship.

Mathematically, Eq. \ref{eqn_heat_exchange_2} can be solved together with Eq. \ref{system_of_equations} by treating  $dT^s(\omega)$ as $dT^{lat}(\omega)$, so that the experimentally obtained AC MCE oscillations will be interpreted as $dT^{\theta}(\omega)$. The heat flow in the real experiment may have a subtle difference from a fully discretized description. Since the size of the thermometer is much smaller than the sample, it cannot exchange heat evenly across the entire sample. Consequently, such consideration has a negligible effect when the sensor is in fairly good thermal contact with the sample. When the thermometer has a limited rate of thermal exchange defined by either geometry or contact materials, the mounting geometry effect starts to affect the thermal transfer behavior of the MCE and cannot be properly described by an ideal discretized thermal model of 2 discretized components.

To experimentally demonstrate the limited thermometer geometry effect above, we designed two experiments for comparison, where the sensor is connected indirectly to the sample.
 The deviation from a single circular feature shown in Fig. \ref{fitresult}(b) is easy to obtain once the thermometer is in loose contact with the measured sample, resulting in an additional characteristic time constant aside from the two described in Eq. \ref{fit_equation_MCE}. Shown in Fig. \ref{extrinsic_mce}(a), when the sensor and the sample are separated by a single quartz component, an additional thermal time constant appears on the lower frequency side, indicated by a deviation of a circular feature on the Cole-Cole plot. Similarly, a third time constant emerges from the higher frequency side when the sample and the sensor are separated by a thermal lead component, such as a thin gold wire in between. In summary, the thermometer effect can be addressed by observing an additional time constant from the measurement by a Cole-Cole analysis. A sample in the absence of spin-lattice relaxation, such as \ce{GdVO4}, will be helpful for experimentalists to calibrate the probe geometry design and optimize the thermal mounting condition. Despite a reduction of the out-of-phase component, the optimized mounting recipe shown in the main text is important to obtain the intrinsic relaxation effect in \ce{YbVO4} that in consistent with the $\chi^{ac}$ measurement.

\end{appendices}

\bibliography{reference}

\begin{thebibliography}{32}%
\makeatletter
\providecommand \@ifxundefined [1]{%
 \@ifx{#1\undefined}
}%
\providecommand \@ifnum [1]{%
 \ifnum #1\expandafter \@firstoftwo
 \else \expandafter \@secondoftwo
 \fi
}%
\providecommand \@ifx [1]{%
 \ifx #1\expandafter \@firstoftwo
 \else \expandafter \@secondoftwo
 \fi
}%
\providecommand \natexlab [1]{#1}%
\providecommand \enquote  [1]{``#1''}%
\providecommand \bibnamefont  [1]{#1}%
\providecommand \bibfnamefont [1]{#1}%
\providecommand \citenamefont [1]{#1}%
\providecommand \href@noop [0]{\@secondoftwo}%
\providecommand \href [0]{\begingroup \@sanitize@url \@href}%
\providecommand \@href[1]{\@@startlink{#1}\@@href}%
\providecommand \@@href[1]{\endgroup#1\@@endlink}%
\providecommand \@sanitize@url [0]{\catcode `\\12\catcode `\$12\catcode `\&12\catcode `\#12\catcode `\^12\catcode `\_12\catcode `\%12\relax}%
\providecommand \@@startlink[1]{}%
\providecommand \@@endlink[0]{}%
\providecommand \url  [0]{\begingroup\@sanitize@url \@url }%
\providecommand \@url [1]{\endgroup\@href {#1}{\urlprefix }}%
\providecommand \urlprefix  [0]{URL }%
\providecommand \Eprint [0]{\href }%
\providecommand \doibase [0]{http://dx.doi.org/}%
\providecommand \selectlanguage [0]{\@gobble}%
\providecommand \bibinfo  [0]{\@secondoftwo}%
\providecommand \bibfield  [0]{\@secondoftwo}%
\providecommand \translation [1]{[#1]}%
\providecommand \BibitemOpen [0]{}%
\providecommand \bibitemStop [0]{}%
\providecommand \bibitemNoStop [0]{.\EOS\space}%
\providecommand \EOS [0]{\spacefactor3000\relax}%
\providecommand \BibitemShut  [1]{\csname bibitem#1\endcsname}%
\let\auto@bib@innerbib\@empty
\bibitem [{\citenamefont {Topping}\ and\ \citenamefont {Blundell}(2018)}]{Topping_2019}%
  \BibitemOpen
  \bibfield  {author} {\bibinfo {author} {\bibfnamefont {C.~V.}\ \bibnamefont {Topping}}\ and\ \bibinfo {author} {\bibfnamefont {S.~J.}\ \bibnamefont {Blundell}},\ }\href {\doibase 10.1088/1361-648X/aaed96} {\bibfield  {journal} {\bibinfo  {journal} {Journal of Physics: Condensed Matter}\ }\textbf {\bibinfo {volume} {31}},\ \bibinfo {pages} {013001} (\bibinfo {year} {2018})}\BibitemShut {NoStop}%
\bibitem [{\citenamefont {Orbach}(1961)}]{Orbach1961}%
  \BibitemOpen
  \bibfield  {author} {\bibinfo {author} {\bibfnamefont {R.}~\bibnamefont {Orbach}},\ }\href {http://www.jstor.org/stable/2414099} {\bibfield  {journal} {\bibinfo  {journal} {Proceedings of the Royal Society of London. Series A, Mathematical and Physical Sciences}\ }\textbf {\bibinfo {volume} {264}},\ \bibinfo {pages} {458} (\bibinfo {year} {1961})}\BibitemShut {NoStop}%
\bibitem [{\citenamefont {Quilliam}\ \emph {et~al.}(2008)\citenamefont {Quilliam}, \citenamefont {Meng}, \citenamefont {Mugford},\ and\ \citenamefont {Kycia}}]{Quilliam2008}%
  \BibitemOpen
  \bibfield  {author} {\bibinfo {author} {\bibfnamefont {J.~A.}\ \bibnamefont {Quilliam}}, \bibinfo {author} {\bibfnamefont {S.}~\bibnamefont {Meng}}, \bibinfo {author} {\bibfnamefont {C.~G.~A.}\ \bibnamefont {Mugford}}, \ and\ \bibinfo {author} {\bibfnamefont {J.~B.}\ \bibnamefont {Kycia}},\ }\href {\doibase 10.1103/PhysRevLett.101.187204} {\bibfield  {journal} {\bibinfo  {journal} {Phys. Rev. Lett.}\ }\textbf {\bibinfo {volume} {101}},\ \bibinfo {pages} {187204} (\bibinfo {year} {2008})}\BibitemShut {NoStop}%
\bibitem [{\citenamefont {Quilliam}\ \emph {et~al.}(2011)\citenamefont {Quilliam}, \citenamefont {Yaraskavitch}, \citenamefont {Dabkowska}, \citenamefont {Gaulin},\ and\ \citenamefont {Kycia}}]{Quilliam2011}%
  \BibitemOpen
  \bibfield  {author} {\bibinfo {author} {\bibfnamefont {J.~A.}\ \bibnamefont {Quilliam}}, \bibinfo {author} {\bibfnamefont {L.~R.}\ \bibnamefont {Yaraskavitch}}, \bibinfo {author} {\bibfnamefont {H.~A.}\ \bibnamefont {Dabkowska}}, \bibinfo {author} {\bibfnamefont {B.~D.}\ \bibnamefont {Gaulin}}, \ and\ \bibinfo {author} {\bibfnamefont {J.~B.}\ \bibnamefont {Kycia}},\ }\href {\doibase 10.1103/PhysRevB.83.094424} {\bibfield  {journal} {\bibinfo  {journal} {Phys. Rev. B}\ }\textbf {\bibinfo {volume} {83}},\ \bibinfo {pages} {094424} (\bibinfo {year} {2011})}\BibitemShut {NoStop}%
\bibitem [{\citenamefont {G{\'{o}}mez-Coca}\ \emph {et~al.}(2014)\citenamefont {G{\'{o}}mez-Coca}, \citenamefont {Urtizberea}, \citenamefont {Cremades}, \citenamefont {Alonso}, \citenamefont {Cam{\'{o}}n}, \citenamefont {Ruiz},\ and\ \citenamefont {Luis}}]{Coca2014}%
  \BibitemOpen
  \bibfield  {author} {\bibinfo {author} {\bibfnamefont {S.}~\bibnamefont {G{\'{o}}mez-Coca}}, \bibinfo {author} {\bibfnamefont {A.}~\bibnamefont {Urtizberea}}, \bibinfo {author} {\bibfnamefont {E.}~\bibnamefont {Cremades}}, \bibinfo {author} {\bibfnamefont {P.~J.}\ \bibnamefont {Alonso}}, \bibinfo {author} {\bibfnamefont {A.}~\bibnamefont {Cam{\'{o}}n}}, \bibinfo {author} {\bibfnamefont {E.}~\bibnamefont {Ruiz}}, \ and\ \bibinfo {author} {\bibfnamefont {F.}~\bibnamefont {Luis}},\ }\href {\doibase 10.1038/ncomms5300} {\bibfield  {journal} {\bibinfo  {journal} {Nature Communications}\ }\textbf {\bibinfo {volume} {5}},\ \bibinfo {pages} {4300} (\bibinfo {year} {2014})}\BibitemShut {NoStop}%
\bibitem [{\citenamefont {Radhakrishna}\ \emph {et~al.}(1981)\citenamefont {Radhakrishna}, \citenamefont {Hammann},\ and\ \citenamefont {Pari}}]{Radhakrishna1981}%
  \BibitemOpen
  \bibfield  {author} {\bibinfo {author} {\bibfnamefont {P.}~\bibnamefont {Radhakrishna}}, \bibinfo {author} {\bibfnamefont {J.}~\bibnamefont {Hammann}}, \ and\ \bibinfo {author} {\bibfnamefont {P.}~\bibnamefont {Pari}},\ }\href {\doibase https://doi.org/10.1016/0304-8853(81)90044-5} {\bibfield  {journal} {\bibinfo  {journal} {Journal of Magnetism and Magnetic Materials}\ }\textbf {\bibinfo {volume} {23}},\ \bibinfo {pages} {254} (\bibinfo {year} {1981})}\BibitemShut {NoStop}%
\bibitem [{\citenamefont {Warburg}(1881)}]{Warburg1881}%
  \BibitemOpen
  \bibfield  {author} {\bibinfo {author} {\bibfnamefont {E.}~\bibnamefont {Warburg}},\ }\href {\doibase https://doi.org/10.1002/andp.18812490510} {\bibfield  {journal} {\bibinfo  {journal} {Annalen der Physik}\ }\textbf {\bibinfo {volume} {249}},\ \bibinfo {pages} {141} (\bibinfo {year} {1881})}\BibitemShut {NoStop}%
\bibitem [{\citenamefont {Kohama}\ \emph {et~al.}(2010)\citenamefont {Kohama}, \citenamefont {Marcenat}, \citenamefont {Klein},\ and\ \citenamefont {Jaime}}]{Kohama2010}%
  \BibitemOpen
  \bibfield  {author} {\bibinfo {author} {\bibfnamefont {Y.}~\bibnamefont {Kohama}}, \bibinfo {author} {\bibfnamefont {C.}~\bibnamefont {Marcenat}}, \bibinfo {author} {\bibfnamefont {T.}~\bibnamefont {Klein}}, \ and\ \bibinfo {author} {\bibfnamefont {M.}~\bibnamefont {Jaime}},\ }\href {\doibase 10.1063/1.3475155} {\bibfield  {journal} {\bibinfo  {journal} {Review of Scientific Instruments}\ }\textbf {\bibinfo {volume} {81}},\ \bibinfo {pages} {104902} (\bibinfo {year} {2010})}\BibitemShut {NoStop}%
\bibitem [{\citenamefont {Brück}(2005)}]{Bruck_2005}%
  \BibitemOpen
  \bibfield  {author} {\bibinfo {author} {\bibfnamefont {E.}~\bibnamefont {Brück}},\ }\href {\doibase 10.1088/0022-3727/38/23/R01} {\bibfield  {journal} {\bibinfo  {journal} {Journal of Physics D: Applied Physics}\ }\textbf {\bibinfo {volume} {38}},\ \bibinfo {pages} {R381} (\bibinfo {year} {2005})}\BibitemShut {NoStop}%
\bibitem [{\citenamefont {Law}\ \emph {et~al.}(2018)\citenamefont {Law}, \citenamefont {Franco}, \citenamefont {Moreno-Ram{\'{i}}rez}, \citenamefont {Conde}, \citenamefont {Karpenkov}, \citenamefont {Radulov}, \citenamefont {Skokov},\ and\ \citenamefont {Gutfleisch}}]{Law2018}%
  \BibitemOpen
  \bibfield  {author} {\bibinfo {author} {\bibfnamefont {J.~Y.}\ \bibnamefont {Law}}, \bibinfo {author} {\bibfnamefont {V.}~\bibnamefont {Franco}}, \bibinfo {author} {\bibfnamefont {L.~M.}\ \bibnamefont {Moreno-Ram{\'{i}}rez}}, \bibinfo {author} {\bibfnamefont {A.}~\bibnamefont {Conde}}, \bibinfo {author} {\bibfnamefont {D.~Y.}\ \bibnamefont {Karpenkov}}, \bibinfo {author} {\bibfnamefont {I.}~\bibnamefont {Radulov}}, \bibinfo {author} {\bibfnamefont {K.~P.}\ \bibnamefont {Skokov}}, \ and\ \bibinfo {author} {\bibfnamefont {O.}~\bibnamefont {Gutfleisch}},\ }\href {\doibase 10.1038/s41467-018-05111-w} {\bibfield  {journal} {\bibinfo  {journal} {Nature Communications}\ }\textbf {\bibinfo {volume} {9}},\ \bibinfo {pages} {2680} (\bibinfo {year} {2018})}\BibitemShut {NoStop}%
\bibitem [{\citenamefont {Pereira}\ \emph {et~al.}(2024)\citenamefont {Pereira}, \citenamefont {Almeida}, \citenamefont {Kiefe}, \citenamefont {Amorim}, \citenamefont {Silva}, \citenamefont {Amaral},\ and\ \citenamefont {Belo}}]{Pereira2024}%
  \BibitemOpen
  \bibfield  {author} {\bibinfo {author} {\bibfnamefont {C.}~\bibnamefont {Pereira}}, \bibinfo {author} {\bibfnamefont {R.}~\bibnamefont {Almeida}}, \bibinfo {author} {\bibfnamefont {R.}~\bibnamefont {Kiefe}}, \bibinfo {author} {\bibfnamefont {C.}~\bibnamefont {Amorim}}, \bibinfo {author} {\bibfnamefont {D.}~\bibnamefont {Silva}}, \bibinfo {author} {\bibfnamefont {J.}~\bibnamefont {Amaral}}, \ and\ \bibinfo {author} {\bibfnamefont {J.}~\bibnamefont {Belo}},\ }\href {\doibase https://doi.org/10.1016/j.jallcom.2023.173290} {\bibfield  {journal} {\bibinfo  {journal} {Journal of Alloys and Compounds}\ }\textbf {\bibinfo {volume} {976}},\ \bibinfo {pages} {173290} (\bibinfo {year} {2024})}\BibitemShut {NoStop}%
\bibitem [{\citenamefont {Fischer}\ \emph {et~al.}(1991)\citenamefont {Fischer}, \citenamefont {Hoffmann}, \citenamefont {Kahle},\ and\ \citenamefont {Paul}}]{FISCHER199179}%
  \BibitemOpen
  \bibfield  {author} {\bibinfo {author} {\bibfnamefont {B.}~\bibnamefont {Fischer}}, \bibinfo {author} {\bibfnamefont {J.}~\bibnamefont {Hoffmann}}, \bibinfo {author} {\bibfnamefont {H.}~\bibnamefont {Kahle}}, \ and\ \bibinfo {author} {\bibfnamefont {W.}~\bibnamefont {Paul}},\ }\href {\doibase https://doi.org/10.1016/0304-8853(91)90115-Q} {\bibfield  {journal} {\bibinfo  {journal} {Journal of Magnetism and Magnetic Materials}\ }\textbf {\bibinfo {volume} {94}},\ \bibinfo {pages} {79} (\bibinfo {year} {1991})}\BibitemShut {NoStop}%
\bibitem [{\citenamefont {Tokiwa}\ and\ \citenamefont {Gegenwart}(2011)}]{Tokiwa2011}%
  \BibitemOpen
  \bibfield  {author} {\bibinfo {author} {\bibfnamefont {Y.}~\bibnamefont {Tokiwa}}\ and\ \bibinfo {author} {\bibfnamefont {P.}~\bibnamefont {Gegenwart}},\ }\href {\doibase 10.1063/1.3529433} {\bibfield  {journal} {\bibinfo  {journal} {Review of Scientific Instruments}\ }\textbf {\bibinfo {volume} {82}},\ \bibinfo {pages} {013905} (\bibinfo {year} {2011})}\BibitemShut {NoStop}%
\bibitem [{\citenamefont {Aliev}\ \emph {et~al.}(2016)\citenamefont {Aliev}, \citenamefont {Batdalov}, \citenamefont {Khanov}, \citenamefont {Koledov}, \citenamefont {Shavrov}, \citenamefont {Tereshina},\ and\ \citenamefont {Taskaev}}]{ALIEV2016601}%
  \BibitemOpen
  \bibfield  {author} {\bibinfo {author} {\bibfnamefont {A.}~\bibnamefont {Aliev}}, \bibinfo {author} {\bibfnamefont {A.}~\bibnamefont {Batdalov}}, \bibinfo {author} {\bibfnamefont {L.}~\bibnamefont {Khanov}}, \bibinfo {author} {\bibfnamefont {V.}~\bibnamefont {Koledov}}, \bibinfo {author} {\bibfnamefont {V.}~\bibnamefont {Shavrov}}, \bibinfo {author} {\bibfnamefont {I.}~\bibnamefont {Tereshina}}, \ and\ \bibinfo {author} {\bibfnamefont {S.}~\bibnamefont {Taskaev}},\ }\href {\doibase https://doi.org/10.1016/j.jallcom.2016.03.238} {\bibfield  {journal} {\bibinfo  {journal} {Journal of Alloys and Compounds}\ }\textbf {\bibinfo {volume} {676}},\ \bibinfo {pages} {601} (\bibinfo {year} {2016})}\BibitemShut {NoStop}%
\bibitem [{ALI(2022)}]{ALIEV2022169300}%
  \BibitemOpen
  \href {\doibase https://doi.org/10.1016/j.jmmm.2022.169300} {\bibfield  {journal} {\bibinfo  {journal} {Journal of Magnetism and Magnetic Materials}\ }\textbf {\bibinfo {volume} {553}},\ \bibinfo {pages} {169300} (\bibinfo {year} {2022})}\BibitemShut {NoStop}%
\bibitem [{\citenamefont {Kazeǐ}\ and\ \citenamefont {Chanieva}(2006)}]{Kaze2006}%
  \BibitemOpen
  \bibfield  {author} {\bibinfo {author} {\bibfnamefont {Z.~A.}\ \bibnamefont {Kazeǐ}}\ and\ \bibinfo {author} {\bibfnamefont {R.~I.}\ \bibnamefont {Chanieva}},\ }\href {https://api.semanticscholar.org/CorpusID:122564464} {\bibfield  {journal} {\bibinfo  {journal} {Journal of Experimental and Theoretical Physics}\ }\textbf {\bibinfo {volume} {102}},\ \bibinfo {pages} {266} (\bibinfo {year} {2006})}\BibitemShut {NoStop}%
\bibitem [{\citenamefont {Palacios}\ \emph {et~al.}(2018)\citenamefont {Palacios}, \citenamefont {Evangelisti}, \citenamefont {S\'aez-Puche}, \citenamefont {Dos Santos-Garc\'{\i}a}, \citenamefont {Fern\'andez-Mart\'{\i}nez}, \citenamefont {Cascales}, \citenamefont {Castro}, \citenamefont {Burriel}, \citenamefont {Fabelo},\ and\ \citenamefont {Rodr\'{\i}guez-Velamaz\'an}}]{Palacios2018}%
  \BibitemOpen
  \bibfield  {author} {\bibinfo {author} {\bibfnamefont {E.}~\bibnamefont {Palacios}}, \bibinfo {author} {\bibfnamefont {M.}~\bibnamefont {Evangelisti}}, \bibinfo {author} {\bibfnamefont {R.}~\bibnamefont {S\'aez-Puche}}, \bibinfo {author} {\bibfnamefont {A.~J.}\ \bibnamefont {Dos Santos-Garc\'{\i}a}}, \bibinfo {author} {\bibfnamefont {F.}~\bibnamefont {Fern\'andez-Mart\'{\i}nez}}, \bibinfo {author} {\bibfnamefont {C.}~\bibnamefont {Cascales}}, \bibinfo {author} {\bibfnamefont {M.}~\bibnamefont {Castro}}, \bibinfo {author} {\bibfnamefont {R.}~\bibnamefont {Burriel}}, \bibinfo {author} {\bibfnamefont {O.}~\bibnamefont {Fabelo}}, \ and\ \bibinfo {author} {\bibfnamefont {J.~A.}\ \bibnamefont {Rodr\'{\i}guez-Velamaz\'an}},\ }\href {\doibase 10.1103/PhysRevB.97.214401} {\bibfield  {journal} {\bibinfo  {journal} {Phys. Rev. B}\ }\textbf {\bibinfo {volume} {97}},\ \bibinfo {pages} {214401} (\bibinfo {year} {2018})}\BibitemShut {NoStop}%
\bibitem [{\citenamefont {Tang}(2024)}]{Tang24}%
  \BibitemOpen
  \bibfield  {author} {\bibinfo {author} {\bibfnamefont {Y.}~\bibnamefont {Tang}},\ }\enquote {\bibinfo {title} {Heat transfer and thermal analysis with computational fluid dynamics},}\ \ (\bibinfo  {publisher} {IntechOpen},\ \bibinfo {address} {Rijeka},\ \bibinfo {year} {2024})\BibitemShut {NoStop}%
\bibitem [{\citenamefont {Iguchi}\ \emph {et~al.}(2023)\citenamefont {Iguchi}, \citenamefont {Fukuda}, \citenamefont {Kano}, \citenamefont {Teranishi},\ and\ \citenamefont {Uchida}}]{Iguchi2023}%
  \BibitemOpen
  \bibfield  {author} {\bibinfo {author} {\bibfnamefont {R.}~\bibnamefont {Iguchi}}, \bibinfo {author} {\bibfnamefont {D.}~\bibnamefont {Fukuda}}, \bibinfo {author} {\bibfnamefont {J.}~\bibnamefont {Kano}}, \bibinfo {author} {\bibfnamefont {T.}~\bibnamefont {Teranishi}}, \ and\ \bibinfo {author} {\bibfnamefont {K.-i.}\ \bibnamefont {Uchida}},\ }\href {\doibase 10.1063/5.0137686} {\bibfield  {journal} {\bibinfo  {journal} {Applied Physics Letters}\ }\textbf {\bibinfo {volume} {122}},\ \bibinfo {pages} {082903} (\bibinfo {year} {2023})}\BibitemShut {NoStop}%
\bibitem [{\citenamefont {Feigelson}(1968)}]{feigelson1968flux}%
  \BibitemOpen
  \bibfield  {author} {\bibinfo {author} {\bibfnamefont {R.}~\bibnamefont {Feigelson}},\ }\href@noop {} {\bibfield  {journal} {\bibinfo  {journal} {Journal of the American Ceramic Society}\ }\textbf {\bibinfo {volume} {51}},\ \bibinfo {pages} {538} (\bibinfo {year} {1968})}\BibitemShut {NoStop}%
\bibitem [{\citenamefont {Smith}\ and\ \citenamefont {Wanklyn}(1974)}]{smith1974flux}%
  \BibitemOpen
  \bibfield  {author} {\bibinfo {author} {\bibfnamefont {S.}~\bibnamefont {Smith}}\ and\ \bibinfo {author} {\bibfnamefont {B.}~\bibnamefont {Wanklyn}},\ }\href@noop {} {\bibfield  {journal} {\bibinfo  {journal} {Journal of Crystal Growth}\ }\textbf {\bibinfo {volume} {21}},\ \bibinfo {pages} {23} (\bibinfo {year} {1974})}\BibitemShut {NoStop}%
\bibitem [{\citenamefont {Oka}\ \emph {et~al.}(2006)\citenamefont {Oka}, \citenamefont {Unoki}, \citenamefont {Shibata},\ and\ \citenamefont {Eisaki}}]{oka2006crystal}%
  \BibitemOpen
  \bibfield  {author} {\bibinfo {author} {\bibfnamefont {K.}~\bibnamefont {Oka}}, \bibinfo {author} {\bibfnamefont {H.}~\bibnamefont {Unoki}}, \bibinfo {author} {\bibfnamefont {H.}~\bibnamefont {Shibata}}, \ and\ \bibinfo {author} {\bibfnamefont {H.}~\bibnamefont {Eisaki}},\ }\href@noop {} {\bibfield  {journal} {\bibinfo  {journal} {Journal of crystal growth}\ }\textbf {\bibinfo {volume} {286}},\ \bibinfo {pages} {288} (\bibinfo {year} {2006})}\BibitemShut {NoStop}%
\bibitem [{\citenamefont {Khansili}\ \emph {et~al.}(2023)\citenamefont {Khansili}, \citenamefont {Bangura}, \citenamefont {McDonald}, \citenamefont {Ramshaw}, \citenamefont {Rydh},\ and\ \citenamefont {Shekhter}}]{Khansili2023}%
  \BibitemOpen
  \bibfield  {author} {\bibinfo {author} {\bibfnamefont {A.}~\bibnamefont {Khansili}}, \bibinfo {author} {\bibfnamefont {A.}~\bibnamefont {Bangura}}, \bibinfo {author} {\bibfnamefont {R.~D.}\ \bibnamefont {McDonald}}, \bibinfo {author} {\bibfnamefont {B.~J.}\ \bibnamefont {Ramshaw}}, \bibinfo {author} {\bibfnamefont {A.}~\bibnamefont {Rydh}}, \ and\ \bibinfo {author} {\bibfnamefont {A.}~\bibnamefont {Shekhter}},\ }\href {\doibase 10.1103/PhysRevB.107.195145} {\bibfield  {journal} {\bibinfo  {journal} {Phys. Rev. B}\ }\textbf {\bibinfo {volume} {107}},\ \bibinfo {pages} {195145} (\bibinfo {year} {2023})}\BibitemShut {NoStop}%
\bibitem [{\citenamefont {Landau}\ and\ \citenamefont {Lifshitz}(1980)}]{Landau}%
  \BibitemOpen
  \bibfield  {author} {\bibinfo {author} {\bibfnamefont {L.~D.}\ \bibnamefont {Landau}}\ and\ \bibinfo {author} {\bibfnamefont {E.~M.}\ \bibnamefont {Lifshitz}},\ }\href@noop {} {\emph {\bibinfo {title} {Statistical Physics, Volume 5, Chapter 110}}}\ (\bibinfo  {publisher} {Butterworth-Heinemann},\ \bibinfo {year} {1980})\BibitemShut {NoStop}%
\bibitem [{Note1()}]{Note1}%
  \BibitemOpen
  \bibinfo {note} {In principle, we can also use $C^{lat}$ to define the $\tau _{ext}$. We eventually chose to use $C^{4f}$ instead of $C^{lat}$ for simplicity during fitting analysis.}\BibitemShut {Stop}%
\bibitem [{\citenamefont {Onsager}(1931)}]{Onsager}%
  \BibitemOpen
  \bibfield  {author} {\bibinfo {author} {\bibfnamefont {L.}~\bibnamefont {Onsager}},\ }\href {\doibase 10.1103/PhysRev.37.405} {\bibfield  {journal} {\bibinfo  {journal} {Phys. Rev.}\ }\textbf {\bibinfo {volume} {37}},\ \bibinfo {pages} {405} (\bibinfo {year} {1931})}\BibitemShut {NoStop}%
\bibitem [{\citenamefont {Nipko}\ \emph {et~al.}(1997)\citenamefont {Nipko}, \citenamefont {Loong}, \citenamefont {Kern}, \citenamefont {Abraham},\ and\ \citenamefont {Boatner}}]{NIPKO1997}%
  \BibitemOpen
  \bibfield  {author} {\bibinfo {author} {\bibfnamefont {J.}~\bibnamefont {Nipko}}, \bibinfo {author} {\bibfnamefont {C.-K.}\ \bibnamefont {Loong}}, \bibinfo {author} {\bibfnamefont {S.}~\bibnamefont {Kern}}, \bibinfo {author} {\bibfnamefont {M.}~\bibnamefont {Abraham}}, \ and\ \bibinfo {author} {\bibfnamefont {L.}~\bibnamefont {Boatner}},\ }\href {\doibase https://doi.org/10.1016/S0925-8388(96)02565-0} {\bibfield  {journal} {\bibinfo  {journal} {Journal of Alloys and Compounds}\ }\textbf {\bibinfo {volume} {250}},\ \bibinfo {pages} {569} (\bibinfo {year} {1997})}\BibitemShut {NoStop}%
\bibitem [{\citenamefont {Santos}\ \emph {et~al.}(2007{\natexlab{a}})\citenamefont {Santos}, \citenamefont {Silva}, \citenamefont {Ayala}, \citenamefont {Guedes}, \citenamefont {Pizani}, \citenamefont {Loong},\ and\ \citenamefont {Boatner}}]{Santos2007}%
  \BibitemOpen
  \bibfield  {author} {\bibinfo {author} {\bibfnamefont {C.~C.}\ \bibnamefont {Santos}}, \bibinfo {author} {\bibfnamefont {E.~N.}\ \bibnamefont {Silva}}, \bibinfo {author} {\bibfnamefont {A.~P.}\ \bibnamefont {Ayala}}, \bibinfo {author} {\bibfnamefont {I.}~\bibnamefont {Guedes}}, \bibinfo {author} {\bibfnamefont {P.~S.}\ \bibnamefont {Pizani}}, \bibinfo {author} {\bibfnamefont {C.-K.}\ \bibnamefont {Loong}}, \ and\ \bibinfo {author} {\bibfnamefont {L.~A.}\ \bibnamefont {Boatner}},\ }\href {\doibase 10.1063/1.2437676} {\bibfield  {journal} {\bibinfo  {journal} {Journal of Applied Physics}\ }\textbf {\bibinfo {volume} {101}},\ \bibinfo {pages} {053511} (\bibinfo {year} {2007}{\natexlab{a}})}\BibitemShut {NoStop}%
\bibitem [{\citenamefont {Santos}\ \emph {et~al.}(2007{\natexlab{b}})\citenamefont {Santos}, \citenamefont {Guedes}, \citenamefont {Loong},\ and\ \citenamefont {Boatner}}]{SANTOS2007_2}%
  \BibitemOpen
  \bibfield  {author} {\bibinfo {author} {\bibfnamefont {C.}~\bibnamefont {Santos}}, \bibinfo {author} {\bibfnamefont {I.}~\bibnamefont {Guedes}}, \bibinfo {author} {\bibfnamefont {C.-K.}\ \bibnamefont {Loong}}, \ and\ \bibinfo {author} {\bibfnamefont {L.}~\bibnamefont {Boatner}},\ }\href {\doibase https://doi.org/10.1016/j.vibspec.2007.05.002} {\bibfield  {journal} {\bibinfo  {journal} {Vibrational Spectroscopy}\ }\textbf {\bibinfo {volume} {45}},\ \bibinfo {pages} {95} (\bibinfo {year} {2007}{\natexlab{b}})},\ \bibinfo {note} {raman Spectroscopy Workshop 2006}\BibitemShut {NoStop}%
\bibitem [{\citenamefont {Mangum}\ and\ \citenamefont {Thornton}(1972)}]{Mangum1972}%
  \BibitemOpen
  \bibfield  {author} {\bibinfo {author} {\bibfnamefont {B.~W.}\ \bibnamefont {Mangum}}\ and\ \bibinfo {author} {\bibfnamefont {D.~D.}\ \bibnamefont {Thornton}},\ }\href {\doibase 10.1063/1.3699447} {\bibfield  {journal} {\bibinfo  {journal} {AIP Conference Proceedings}\ }\textbf {\bibinfo {volume} {5}},\ \bibinfo {pages} {311} (\bibinfo {year} {1972})}\BibitemShut {NoStop}%
\bibitem [{\citenamefont {Massat}\ \emph {et~al.}(2022)\citenamefont {Massat}, \citenamefont {Wen}, \citenamefont {Jiang}, \citenamefont {Hristov}, \citenamefont {Liu}, \citenamefont {Smaha}, \citenamefont {Feigelson}, \citenamefont {Lee}, \citenamefont {Fernandes},\ and\ \citenamefont {Fisher}}]{tmvo42022}%
  \BibitemOpen
  \bibfield  {author} {\bibinfo {author} {\bibfnamefont {P.}~\bibnamefont {Massat}}, \bibinfo {author} {\bibfnamefont {J.}~\bibnamefont {Wen}}, \bibinfo {author} {\bibfnamefont {J.~M.}\ \bibnamefont {Jiang}}, \bibinfo {author} {\bibfnamefont {A.~T.}\ \bibnamefont {Hristov}}, \bibinfo {author} {\bibfnamefont {Y.}~\bibnamefont {Liu}}, \bibinfo {author} {\bibfnamefont {R.~W.}\ \bibnamefont {Smaha}}, \bibinfo {author} {\bibfnamefont {R.~S.}\ \bibnamefont {Feigelson}}, \bibinfo {author} {\bibfnamefont {Y.~S.}\ \bibnamefont {Lee}}, \bibinfo {author} {\bibfnamefont {R.~M.}\ \bibnamefont {Fernandes}}, \ and\ \bibinfo {author} {\bibfnamefont {I.~R.}\ \bibnamefont {Fisher}},\ }\href {https://www.pnas.org/doi/full/10.1073/pnas.2119942119} {\bibfield  {journal} {\bibinfo  {journal} {Proceedings of the National Academy of Sciences}\ }\textbf {\bibinfo {volume} {119}},\ \bibinfo {pages} {e2119942119} (\bibinfo {year} {2022})}\BibitemShut {NoStop}%
\bibitem [{\citenamefont {Vallipuram}\ \emph {et~al.}(2024)\citenamefont {Vallipuram}, \citenamefont {Chen}, \citenamefont {Campillo}, \citenamefont {Mezidi}, \citenamefont {Grissonnanche}, \citenamefont {Boulanger}, \citenamefont {Lefran\ifmmode~\mbox{\c{c}}\else \c{c}\fi{}ois}, \citenamefont {Zic}, \citenamefont {Li}, \citenamefont {Fisher}, \citenamefont {Baglo},\ and\ \citenamefont {Taillefer}}]{Vallipuram2024}%
  \BibitemOpen
  \bibfield  {author} {\bibinfo {author} {\bibfnamefont {A.}~\bibnamefont {Vallipuram}}, \bibinfo {author} {\bibfnamefont {L.}~\bibnamefont {Chen}}, \bibinfo {author} {\bibfnamefont {E.}~\bibnamefont {Campillo}}, \bibinfo {author} {\bibfnamefont {M.}~\bibnamefont {Mezidi}}, \bibinfo {author} {\bibfnamefont {G.}~\bibnamefont {Grissonnanche}}, \bibinfo {author} {\bibfnamefont {M.-E.}\ \bibnamefont {Boulanger}}, \bibinfo {author} {\bibfnamefont {E.}~\bibnamefont {Lefran\ifmmode~\mbox{\c{c}}\else \c{c}\fi{}ois}}, \bibinfo {author} {\bibfnamefont {M.~P.}\ \bibnamefont {Zic}}, \bibinfo {author} {\bibfnamefont {Y.}~\bibnamefont {Li}}, \bibinfo {author} {\bibfnamefont {I.~R.}\ \bibnamefont {Fisher}}, \bibinfo {author} {\bibfnamefont {J.}~\bibnamefont {Baglo}}, \ and\ \bibinfo {author} {\bibfnamefont {L.}~\bibnamefont {Taillefer}},\ }\href {\doibase 10.1103/PhysRevB.110.045144} {\bibfield  {journal} {\bibinfo  {journal} {Phys. Rev. B}\ }\textbf {\bibinfo {volume} {110}},\ \bibinfo {pages} {045144} (\bibinfo {year}
  {2024})}\BibitemShut {NoStop}%
\end{thebibliography}%

\end{document}